\documentclass[12pt]{article}
\usepackage{epsfig}
\usepackage{amssymb}
\usepackage{amsmath}
\usepackage{amsfonts}
\usepackage{mathrsfs}
\DeclareMathAlphabet{\mathscrbf}{OMS}{mdugm}{b}{n}
\usepackage[dvips]{color}
\usepackage{multirow}
\usepackage{calc}
\usepackage{accents}

\makeatletter
\newcommand{\shorteq}{\mathrel{\mkern0.2mu\mathpalette\shorteq@\relax\mkern0.2mu}}
\newcommand{\shorteq@}[2]{\scalebox{0.5}[1]{$\m@th#1=$}}

\newcommand{\longeq}[1]{\mathrel{\mathpalette\longeq@{#1}}}
\newcommand{\longeq@}[2]{%
  \begingroup
  \sbox\z@{$\m@th#1=$}%
  \ifdim#2<\wd\z@
    \resizebox{#2}{\height}{\box\z@}%
  \else
    \ifdim#2<3\wd\z@
      \hbox to #2{$\m@th#1=\hss=\hss=\hss=$}%
    \else
      \hbox to #2{$\m@th#1=\cleaders\hbox to 0.2\wd\z@{\hss$#1=$\hss}\hfil=$}%
    \fi
  \fi
  \endgroup
}
\makeatother

\newcommand{\bsigma}{\boldsymbol{\sigma}}

\newcommand{\bnabla}{\boldsymbol{\nabla}}

\newcommand{\R}{\mathbb{R}}
\newcommand{\C}{\mathbb{C}}

\newcommand{\fa}{\mathfrak{a}}

\newcommand{\fc}{\mathfrak{c}}

\newcommand{\ff}{\mathfrak{f}}

\newcommand{\fn}{{\mathfrak{n}}}

\newcommand{\fs}{\mathfrak{s}}

\newcommand{\fz}{\mathfrak{z}}

\newcommand{\fK}{\mathfrak{K}}

\newcommand{\bfe}{\mathbf{e}}

\newcommand{\bk}{\mathbf{k}}

\newcommand{\bfr}{\mathbf{r}}

\newcommand{\bx}{\mathbf{x}}

\newcommand{\bA}{\mathbf{A}}

\newcommand{\bB}{\mathbf{B}}

\newcommand{\bcE}{{\boldsymbol{\cE}}}
\newcommand{\bF}{\mathbf{F}}
\newcommand{\bG}{\mathbf{G}}
\newcommand{\bH}{\mathbf{H}}
\newcommand{\bI}{\mathbf{I}}

\newcommand{\bL}{\mathbf{L}}

\newcommand{\bM}{\mathbf{M}}

\newcommand{\bP}{\mathbf{P}}
\newcommand{\bQ}{\mathbf{Q}}

\newcommand{\bU}{\mathbf{U}}
\newcommand{\bV}{\mathbf{V}}

\newcommand{\cH}{\mathcal{H}}
\newcommand{\cE}{\mathcal{E}}

\newcommand{\cK}{\mathcal{K}}

\newcommand{\cP}{\mathcal{P}}

\newcommand{\cS}{\mathcal{S}}
\newcommand{\cT}{\mathcal{T}}
\newcommand{\cU}{\mathcal{U}}

\newcommand{\be}{\begin{equation}}
\newcommand{\ee}{\end{equation}}
\newcommand{\bea}{\begin{eqnarray}}
\newcommand{\eea}{\end{eqnarray}}
\newcommand{\nn}{\nonumber}
\newcommand{\kt}{\rangle}
\newcommand{\br}{\langle}

\newcommand{\ed}{\end{document}}

\newcommand{\bi}{\begin{itemize}}
\newcommand{\ei}{\end{itemize}}

\newcommand{\bce}{\begin{center}}
\newcommand{\ece}{\end{center}}

\newcommand{\sE}{\mathscr{E}}
\newcommand{\sF}{\mathscr{F}}

\newcommand{\bsL}{\mathscrbf{L}}

\newcommand{\sL}{\mathscr{L}}
\newcommand{\sM}{\mathscr{M}}

\newcommand{\sT}{\mathscr{T}}
\newcommand{\sV}{\mathscr{V}}
\newcommand{\sW}{\mathscr{W}}

\newcommand{\RE}{{\rm Re}}
\newcommand{\IM}{{\rm Im}}

\newcommand{\bcK}{{\boldsymbol{\cK}}}

\newcommand{\bcH}{{\boldsymbol{\cH}}}
\newcommand{\bcU}{{\boldsymbol{\cU}}}

\newcommand{\bzero}{{\boldsymbol{0}}}

\newcommand{\for}{{\mbox{\rm for}}}

\newcommand{\cx}{{\check{x}}}

\begin{document}

\title{Scattering of TE and TM waves by inhomogeneities of a 2D 
material, low-frequency behavior of the scattering amplitude, and  low-frequency invisibility}

\author{Farhang Loran\thanks{E-mail address: loran@iut.ac.ir}
~and Ali~Mostafazadeh\thanks{Corresponding author, E-mail address:
amostafazadeh@ku.edu.tr}\\[6pt]
$^*$Department of Physics, Isfahan University of Technology, \\ Isfahan 84156-83111, Iran\\[6pt]
$^\dagger$Departments of Mathematics and Physics, Ko\c{c} University,\\  34450 Sar{\i}yer,
Istanbul, T\"urkiye}

\date{ }
\maketitle

\begin{abstract} 
The propagation of the transverse electric (TE) and transverse magnetic (TM) waves in an effectively two-dimensional (2D) isotropic medium is described by Bergmann's equation of acoustics. We develop a dynamical formulation of the stationary scattering of these waves and explore its application in the study of the low-frequency behavior of the scattering data. Specifically, we introduce a suitable notion of fundamental transfer matrix for TE and TM waves in 2D. This is an integral operator $\widehat\bM$ that carries the information about the scattering properties of the medium and admits a Dyson series expansion involving a non-Hermitian Hamiltonian operator. For situations where the inhomogeneities of the medium are confined to a layer of thickness $\ell$, we use the Dyson series for $\widehat\bM$ to construct the series expansion of the scattering amplitude in powers of $k\ell$, where $k$ is the incident wavenumber. We derive analytic expressions for the leading- and next-to-leading-order terms of this series, verify the effectiveness of their application to a class of exactly solvable models, and use them to study low-frequency invisibility. In particular, we develop a low-frequency cloaking scheme which is applicable for both TE and TM waves. Our results have immediate applications in the study of low-frequency scattering of acoustic waves in a  2D fluid as these waves are also described by Bergmann's equation.
\end{abstract}

\section{Introduction}

Transfer matrices \cite{sanchez,tjp-2020} have been used as a powerful tool for performing scattering calculations since the 1940's \cite{jones-1941,abeles,thompson,berreman-1972}. The effectiveness of this tool stems from the composition property of transfer matrix which allows for dissecting a scatterer into slices and assembling the contributions of the latter to determine the scattering properties of the scatterer \cite{tjp-2020,McLean-Pendry,Penry-Bell}. For waves propagating in an effectively one-dimensional locally-periodic multilayer medium, this property reduces the solution of the scattering problem to the one for a single period \cite{yeh,pereyra,griffiths}. 

The observation that the composition property of the transfer matrix of potential scattering in one dimension coincides with that of the evolution operators in quantum mechanics \cite{ap-2014} motivated the development of a dynamical formulation of stationary scattering (DFSS) for scalar waves satisfying the Schr\"odinger or Helmholtz equations in two and three dimensions \cite{pra-2021} and electromagnetic waves in three dimensions \cite{pra-2023}. This involves the identification of an effective non-unitary quantum system whose evolution operator determines an integral operator, called the fundamental transfer matrix, and extracting the scattering data from the latter \cite{pra-2021,pra-2023}. Among the triumphs of this approach are a consistent singularity-free treatment of delta-function point scatterers in two and three dimensions \cite{pra-2023,ap-2022} and the discovery of the scattering potentials and permittivity-permeability profiles for which the Born approximation gives the exact solution of the scattering problem \cite{ptep-2024a,jpa-2024}.

Another interesting application of DFSS is the construction of low-frequency expansions of the scattering amplitude for the potential scattering in one dimension \cite{jmp-2021} and the scattering problems defined by the Helmholtz equation in one, two, and three dimensions \cite{jpa-2021,pra-2025}. This, in particular, applies to low-frequency transverse electric (TE) waves scattered by an effectively two-dimensional (2D) nonmagnetic isotropic medium. The purpose of the present article is to extend the approach of Ref.~\cite{pra-2025} to transverse magnetic (TM) waves  \cite{zhuch-1994,yu-1998,kashihara-2006,lechleiter-2014} and TE waves propagating in a general (possibly magnetic) isotropic 2D medium. Because these waves are not described by the Helmholtz equation, this requires the development of a DFSS for them. The first step in this direction has been taken in Ref.~\cite{ptep-2024b} where a DFSS for TE and TM waves propagating in an effectively one-dimensional (1D) isotropic medium is proposed. The results proved to be instrumental in the study of low-frequency scattering of these waves \cite{jpa-2025}. This together with the lack of analytic treatments of low-frequency scattering of electromagnetic waves in 2D materials provide the basic motivation for the present investigation. 

Consider time-harmonic electromagnetic waves propagating in a stationary linear isotropic medium void of free charges and currents that is immersed in a homogeneous background (or vacuum). Let $\varepsilon$ and $\mu$ denote the permittivity and permeability of the medium, $\varepsilon_{\rm b}$ and $\mu_{\rm b}$ be  the permittivity and permeability of the background, and $\hat\varepsilon:=\varepsilon/\varepsilon_{\rm b}$ and $\hat\mu:=\mu/\mu_{\rm b}$ be the relative permittivity and permeability of the medium, respectively. Then, the electric and magnetic fields for such a wave have respectively the forms, $e^{-i\omega t}\bcE(\bfr)/\sqrt\varepsilon_{\rm b}$ and $e^{-i\omega t}\bcH(\bfr)/\sqrt\mu_{\rm b}$, where $\omega$ is the angular frequency of the wave,  $\bcE$ and $\bcH$ are vector-valued functions which in view of Maxwell's equations satisfy
 	\begin{align}
	&\bnabla\cdot(\hat\varepsilon\,\bcE)=0,
	&& \bnabla\cdot(\hat\mu\,\bcH)=0,
	 \label{Coulomb}\\
	 &\bnabla\times\bcE=ik\hat\mu\,\bcH,
	 &&\bnabla\times\bcH=-ik\hat\varepsilon\,\bcE,
    	\label{Amper} 
	\end{align}
$k:=\omega/c_{\rm b}$, and $c_{\rm b}:=1/\sqrt{\varepsilon_{\rm b}\mu_{\rm b}}$.\footnote{If the medium is placed in vacuum, $\varepsilon_{\rm b}$ and $\mu_{\rm b}$ are respectively the permittivity and permeability of vacuum, and $c_{\rm b}$ is the speed of light in vacuum.}

Suppose that the medium has translational symmetry along the $z$ axis of a Cartesian coordinate system, so that $\hat\varepsilon$ and $\hat\mu$ are functions of $x$ and $y$. Then, we can speak of TE and TM waves. {We define these as the electromagnetic waves whose electric and magnetic fields are respectively orthogonal to the plane of incidence \cite{born-wolf}.\footnote{{Our definition of TE and TM waves differs from those adopted in texts in which TE and TM waves are respectively defined as waves whose electric and magnetic fields are orthogonal to the symmetry axis of the medium \cite{taflov}. In particular, what we call TE and TM waves correspond to the ${\rm TM}$ and ${\rm TE}$ waves of Ref.~\cite{taflov}, respectively.}} Taking the latter to be the $x$-$y$ plane, we can identify TE and TM waves with the solutions of \eqref{Coulomb} and \eqref{Amper} that satisfy} 
	\begin{align}
	\bcE(\bfr)&=\psi(x,y)\,\hat\bfe_z~~~\for~~~\mbox{TE waves},
	\label{TE-def}\\
	\bcH(\bfr)&=\psi(x,y)\,\hat\bfe_z~~~\for~~~\mbox{TM waves},
	\label{TM-def}
	\end{align}	
where $\psi:\R^2\to\C$ is a scalar function, and $\hat\bfe_u$ denotes the unit vector along the positive $u$ axis with $u\in\{x,y,z\}$. For the ansatz \eqref{TE-def} and \eqref{TM-def}, Eqs.~\eqref{Coulomb} and \eqref{Amper} imply
	\begin{align}
	&\left.\begin{aligned}
	&\cH_z=\partial_z\cH_x=\partial_z\cH_y=0\\
	&\partial_x\cH_y-\partial_y\cH_x=-ik\,\hat\varepsilon\,\psi
	\end{aligned}\right\}~~\mbox{for TE waves},
	\label{TE-def-2}\\[6pt]
	&\left.\begin{aligned}
	&\cE_z=\partial_z\cE_x=\partial_z\cE_y=0\\
	&\partial_x\cE_y-\partial_y\cE_x=ik\,\hat\mu\,\psi
	\end{aligned}~\right\}~~~\mbox{for TM waves},
	\label{TM-de2}\\[6pt]
	&\vec\nabla\cdot(\alpha^{-1} \vec\nabla \psi)+k^2\beta\,\psi=0,
	\label{Bergmann}
	\end{align}
where $\cE_u$ and $\cH_u$ respectively stand for the $u$ component of $\bcE$ and $\bcH$, $\vec\nabla=\hat\bfe_x\partial_x+\hat\bfe_y\partial_y$, and
	\begin{align}
	&\alpha:=\left\{\begin{array}{ccc}
	\hat\mu&\for&\mbox{TE waves},\\
	\hat\varepsilon&\for&\mbox{TM waves},
	\end{array}\right.
	&&\beta:=\left\{\begin{array}{ccc}
	\hat\varepsilon&\for&\mbox{TE waves},\\
	\hat\mu&\for&\mbox{TM waves}.
	\end{array}\right.
	\label{alpha-beta}
	\end{align}

Multiplying both sides of \eqref{Bergmann} by $\alpha$ and denoting the (complex) refractive index of the medium by $\fn$, which satisfies $\fn^2=\hat\varepsilon\,\hat\mu$, we can identify \eqref{Bergmann} with the Bergmann equation in 2D \cite{Bergmann,Martin}:
	\be
	\alpha\,\vec\nabla\cdot(\alpha^{-1} \vec\nabla \psi)+k^2\fn^2\psi=0.
	\nn
	\ee
This is the wave equation that describes the propagation of TE and TM waves in an isotropic 2D medium \cite{Reutskiy}. The same equation governs the behavior of acoustic waves in a 2D fluid, if we reinterpret $\alpha$ and $\fn$ as the mass density of the fluid and the ratio of the speed of sound at spatial infinity to that in the fluid, respectively \cite{Martin}.

To describe the scattering of TE and TM waves by the inhomogeneities of a 2D isotropic medium, we suppose that the inhomogeneities cause short-range interactions \cite{yafaev} so that bounded solutions of \eqref{Bergmann} tend to the superpositions of plane waves at spatial infinities. In particular, this equation admits scattering solutions satisfying
	\be
	\psi(\vec r)\to
	\frac{1}{2\pi}\left[e^{i\vec k_0\cdot\vec r}+\sqrt{\frac{i}{ k r}}\,e^{ikr}\,\ff(\theta)\right]~~~\for~~~r\to\infty,
	\label{scattering}
	\ee
where $\vec r:=x\,\hat\bfe_x+y\,\hat\bfe_y=(x,y)$, $\vec k_0$ is the incident wave vector, $r$ and $\theta$ are polar coordinates of $\vec r$, and $\ff$ is the scattering amplitude \cite{adhikari-86}.

The outline of the remainder of this article is as follows. In Sec.~\ref{S2}, we provide a quantitative description of the conditions on $\hat\varepsilon$ and $\hat\mu$ under which Bergmann's equation~(\ref{Bergmann}) defines a well-posed scattering problem, introduce an analog of the fundamental transfer matrix of Ref.~\cite{pra-2021} for the scattering problem defined by this equation, and show that it admits a Dyson series expansion determined by an effective non-Hermitian Hamiltonian operator. In Sec.~\ref{S3}, we consider situations where the inhomogeneities of the medium are confined to a strip of thickness $\ell$ and use the Dyson series for the fundamental transfer matrix to expand the scattering amplitude in powers of $k\ell$. Truncating this series gives rise to a hierarchy of low-frequency approximations. In Sec.~\ref{S4}, we apply our general results to a class of exactly solvable models to provide a check on the validity of these approximations. In Sec.~\ref{S5} we derive conditions for low-frequency invisibility and outline a cloaking scheme that prevents the inhomogeneities of the medium to scatter TE and TM waves at low-frequencies regardless of their incidence angles. In Sec.~\ref{S6}, we summarize our findings and present our concluding remarks.

\section{Dynamical formulation of the scattering of TE and TM waves in 2D}
\label{S2}

\subsection{Fundamental transfer matrix and the scattering amplitude}

Suppose that there is a positive real number $a$ such that $\hat\varepsilon(\vec r)$ and $\hat\mu(\vec r)$ have continuous second-order partial derivatives for $r\geq a$. Then, for $r\geq a$, we can use the change of variable, $\psi\to \phi:=\alpha^{-1/2}\psi$, to map \eqref{Bergmann} to the Schr\"odinger equation, 
	\be
	-\nabla^2 \phi+v\,\phi=k^2 \phi,
	\label{sch-eq}
	\ee
where $v:=k^2(1-\alpha\beta)-\alpha ^{1/2}\,\nabla^2\alpha^{-1/2}$. In view of \eqref{alpha-beta}, this is a short-range potential \cite{yafaev}, i.e., it decays faster than $r^{-3/2}$ as $r\to\infty$, if the same holds for 
$\hat\varepsilon(\vec r)-1$,  $\hat\mu(\vec r)-1$,  $\nabla^2\hat\varepsilon(\vec r)$, and $\nabla^2\hat\mu(\vec r)$.
%	\begin{align}
%	&\hat\varepsilon(\vec r)-1, && \hat\mu(\vec r)-1, &&\nabla^2\hat\varepsilon(\vec r), &&\nabla^2\hat\mu(\vec r).
%	\nn%\label{condi-1}
%	\end{align}
Under this condition, the bounded solutions of \eqref{sch-eq} tend to superpositions of plane waves for $r\to\infty$, and this equation admits scattering solutions \cite{yafaev}. Because this condition also implies that $\lim_{r\to\infty}[\psi(\vec r)-\phi(\vec r)]=0$, the same holds for Bergmann's equation \eqref{Bergmann}. This means that for each bounded solution $\psi$ of \eqref{Bergmann}, there are functions (tempered distribution) $A_\pm,B_\pm:\R\to\C$ fulfilling 
	\be
	A_\pm(p)=B_\pm(p)=0~~~\for~~~|p|\geq k,
	\label{condi-0}
	\ee
such that
	\be
	\psi(x,y)\to\int_{-k}^k \frac{dp}{4\pi^2\varpi(p)}
	\Big[A_\pm(p) e^{i\varpi(p)x}+B_\pm(p) e^{-i\varpi(p)x}\Big]e^{ipy}~~~~\for~~~~x\to\pm\infty,
	\label{asym}
	\ee
where %$\varpi:\R\to\C$ is the function defined by
	\be	
	\varpi(p):=\left\{\begin{array}{ccc}
	\sqrt{k^2-p^2}&\for&|p|<k,\\
	i\sqrt{p^2-k^2}&\for&|p|\geq k.\end{array}\right.
	\label{varpi-def}
	\ee
	
Let $\C^{m\times n}$ be the set of $m\times n$ complex matrices, for each positive integer $d$, $\sF^d$ denote the set of functions, $\bF:\R\to\C^{d\times 1}$, and 
	\[\sF^d_k:=\{\:\bF\in\sF^d\,|\,\bF(p)=\bzero~\for~|p|\geq k\:\},\]
where $\bzero$ stands for the zero (null) matrices of all sizes. Then \eqref{condi-0} means $A_\pm,B_\pm\in\sF_k^2$, and in analogy with the treatment of potential scattering in 2D that is given in Ref.~\cite{pra-2021}, we can introduce the fundamental transfer matrix for the medium as the linear operator $\widehat\bM:\sF_k^2\to\sF_k^2$ satisfying
	\be
	\widehat\bM
	\left[\begin{array}{c}
	A_-\\
	B_-\end{array}\right]=\left[\begin{array}{c}
	A_+\\
	B_+\end{array}\right].
	\label{M-def}
	\ee
Applying this equation for scattering solutions of \eqref{Bergmann}, we find \cite{pra-2021}
	\begin{align}
	\ff(\theta)=\frac{-i}{\sqrt{2\pi}}\times\left\{
	\begin{array}{ccc}
	A^l_+(k\sin\theta)-2\pi\delta(\theta-\theta_0)&\for&
	\cos\theta_0>0~\&~\cos\theta>0,\\
	B^l_-(k\sin\theta)&\for&
	\cos\theta_0>0~\&~\cos\theta<0,\\
	A^r_+(k\sin\theta)&\for&
	\cos\theta_0<0~\&~\cos\theta>0,\\
	B^r_-(k\sin\theta)-2\pi\delta(\theta-\theta_0)&\for&
	\cos\theta_0<0~\&~\cos\theta<0,
	\end{array}\right.
	\label{f=}
	\end{align}
where $\theta_0$ is the incidence angle, which satisfies $\cos\theta_0=k^{-1}\hat\bfe_x\cdot\vec\bk_0$,   
	\begin{align}
	&A^l_+:=\widehat M_{12}B^l_-+\widehat M_{11}\check\delta_{p_0},
	&&A^r_+:=\widehat M_{12}B^r_-,
	\label{As=}
	\end{align}
{$\widehat M_{ab}$ denote the entries of $\widehat\bM$,} $\check\delta_{p_0}$ is defined by
	\begin{align}
	&\check\delta_{p_0}(p):=2\pi\varpi(p_0)\:\delta(p-p_0),
	&&p_0:=\hat\bfe_y\cdot\vec\bk_0=k\sin\theta_0,
	\label{tdelta}
	\end{align}
$\delta(\cdot)$ stands for the Dirac delta function, and $B^{l/r}_-$ are the elements of $\sF^1_k$ that solve the following linear equations.
	\begin{align}
	&\widehat M_{22} B^l_-=-\widehat M_{21}\check\delta_{p_0},
	&&\widehat M_{22} B^r_-=\check\delta_{p_0}.
	\label{Bs=}
	\end{align}
		
As noted in Ref.~\cite{pra-2025}, Eqs.~\eqref{Bs=} admit series solutions of the form
	\begin{align}
	&B^l_-=\sum_{j=0}^\infty\widehat N_{22}^j \widehat N_{21}\check\delta_{p_0},
	&&B^r_-=\sum_{j=0}^\infty\widehat N_{22}^j \check\delta_{p_0},
	\label{B-series}
	\end{align}
where %$\widehat N_{ab}:\sF^1_k\to\sF^1_k$ are the linear operators given by 
	\be
	\widehat N_{ab}:=\delta_{ab}\widehat I-\widehat M_{ab},
	\label{N-ab=}
	\ee
and $\delta_{ab}$ denotes the Kronecher delta symbol. Substituting \eqref{B-series} in \eqref{As=}, we find
	\begin{align}
	&A^l_+=\check\delta_{p_0}-\widehat N_{11}\check\delta_{p_0}-
	\sum_{j=0}^\infty
	\widehat N_{12} \widehat N_{22}^j\widehat N_{21} \check\delta_{p_0},
	&&A^r_+=-\sum_{j=0}^\infty\widehat N_{12}\widehat N_{22}^j \check\delta_{p_0}.
	\label{A-series}
	\end{align} 
	
Let $\widehat L$ be a linear operator acting in $\sF^1_k$. Then, we can use Dirac's bra-ket notation and \eqref{tdelta}, to show that
	\begin{align}
	&\check\delta_{p_0}=2\pi \varpi(p_0)|p_0\kt, 
	&&\big(\widehat L\check\delta_{p_0}\big)(p)=2\pi\varpi(p_0)\br p|\widehat L|p_0\kt.
	\label{id-1}
	\end{align}
Employing these identities in \eqref{B-series} and \eqref{A-series}, we find 
	\begin{align}
	&B^l_-(k\sin\theta)=2\pi k|\cos\theta_0|
	\sum_{j=0}^\infty\br p_1|\widehat N_{22}^j \widehat N_{21}|p_0\kt,
	\label{B-L-series} \\
	&B^r_-(k\sin\theta)-2\pi\delta(\theta-\theta_0)=2\pi k|\cos\theta_0|
	\sum_{j=1}^\infty\br p_1|\widehat N_{22}^j |p_0\kt,	
	%=2\pi \varpi(p_0)\sum_{j=0}^\infty\br p|\widehat N_{22}^j |p_0\kt ,
	\label{B-R-series} \\
	&A^l_+(k\sin\theta)-2\pi\delta(\theta-\theta_0)=
	-2\pi k|\cos\theta_0|\Big[
	\br p_1|\widehat N_{11}|p_0\kt+
	\sum_{j=0}^\infty
	\br p_1|\widehat N_{12} \widehat N_{22}^j\widehat N_{21}|p_0\kt\Big],
	\label{A-L-series} \\
	&A^r_+(k\sin\theta)=-2\pi k|\cos\theta_0|\sum_{j=0}^\infty\br p_1|\widehat N_{12}\widehat N_{22}^j |p_0\kt,
	\label{A-R-series} 
	\end{align}
where $p_1:=k\sin\theta$, and we have made use of the relations, $\varpi(p_0)=k|\cos\theta_0|$ and $\check\delta_{p_0}(p_1)=2\pi \delta(\theta-\theta_0)$, which follow from \eqref{varpi-def} and \eqref{tdelta}. Substituting \eqref{B-L-series} -- \eqref{A-R-series} in \eqref{f=} and taking note of \eqref{N-ab=}, we can express the scattering amplitude in terms of the entries of the fundamental transfer matrix. 

\begin{figure}
        \begin{center}
        \includegraphics[scale=.24]{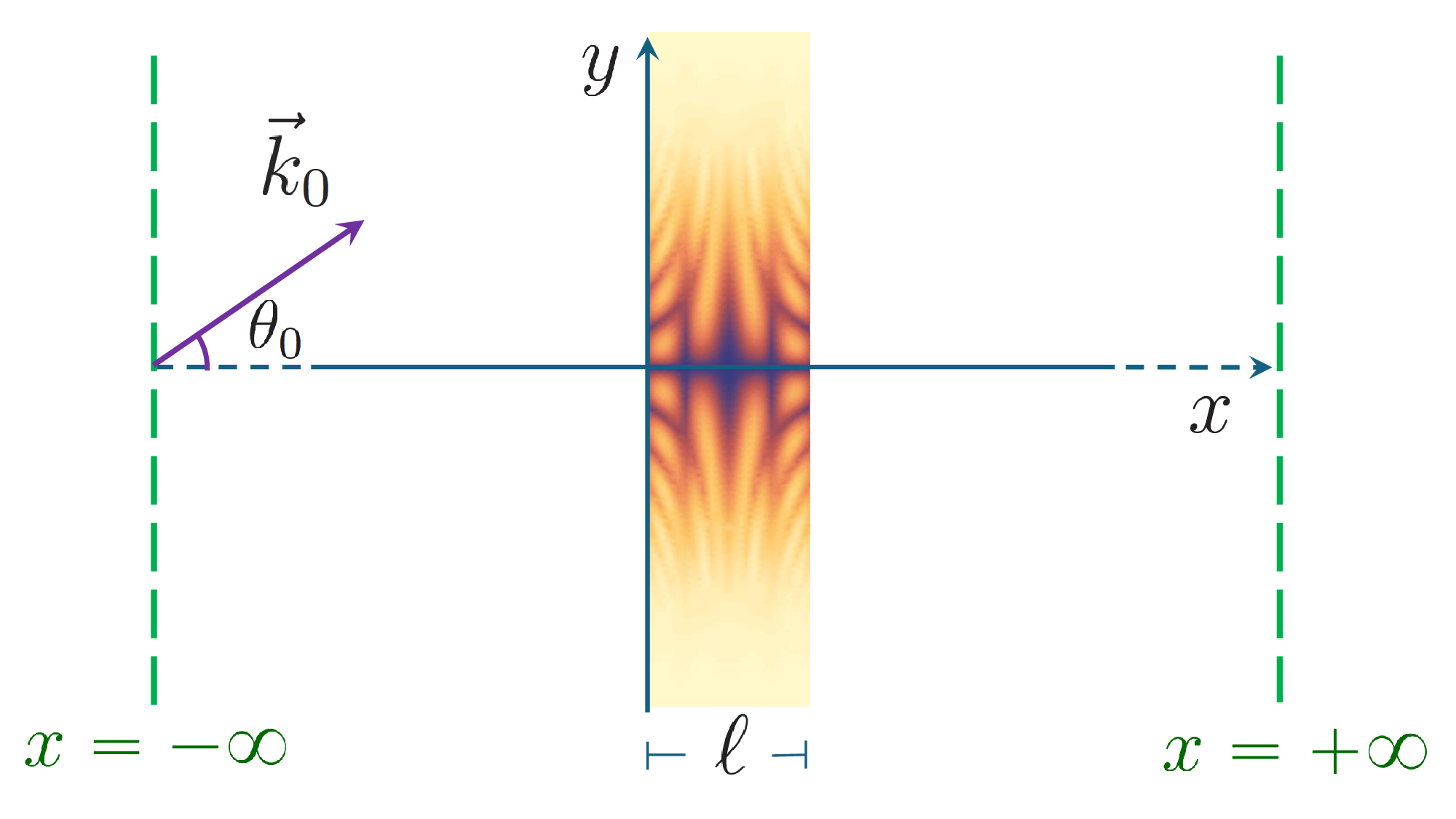}~~~~ 
        \includegraphics[scale=.24]{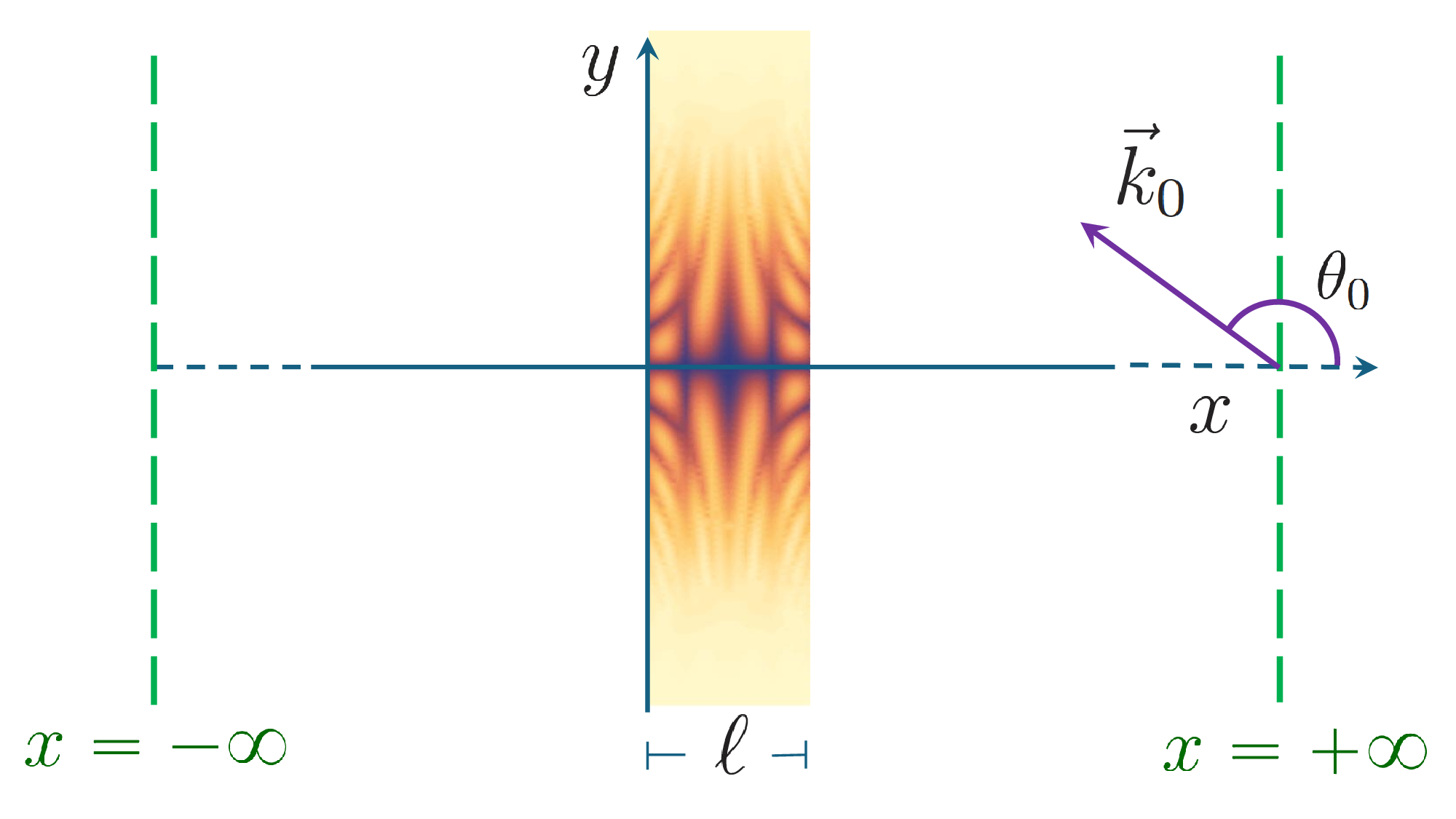} 
        \caption{Schematic views of the scattering setup for the scattering of TE and TM waves by the inhomogeneities of a {generic} effectively two-dimensional isotropic medium that are confined to the region given by $0\leq x\leq\ell$. {The coloring of this region in different shades of brown in a yellow background represents the inhomogeneities of the medium.} The left and right panels correspond to the incident waves with incidence angle $\theta_0$ satisfying $\cos\theta_0>0$ and $\cos\theta_0<0$, respectively. {Dashed green lines represent the planes given by $x=\pm\infty$ where the detectors are located.}}
        \label{fig1}
        \end{center}
        \end{figure}

{The dynamical formulation of the stationary scattering of TE and TM waves that we have outlined in this section corresponds to a scattering setup in which the source of the incident wave is placed on either of the planes $x=-\infty$ or $x=+\infty$, and the detectors observing the scattered wave are mounted on both of these planes \cite{pra-2021}. This is reflected in the fact that \eqref{f=} gives the scattering amplitude for all values of $\theta$ and $\theta_0$ except for $\theta_0,\theta\in\{90^\circ,270^\circ\}$. Because the scattering amplitude for a short-range interaction is a continuous function of $\theta$ and $\theta_0$, this does not lead to any loss of generality. Note also that this scattering setup is particularly natural for situations where the inhomogeneities of the medium causing the scattering are confined to a region bounded by a pair of planes that are orthogonal to the $x$-axis, as depicted in Fig.~\ref{fig1}.}    

\subsection{Dyson series expansion of the fundamental transfer matrix}

%Similarly to the case of potential scattering \cite{pra-2021}, the fundamental transfer matrix associated with the scattering of TE and TM waves in 2D is given by the evolution operator for an effective non-unitary quantum system. To first step toward the identification of this system is to express Bergmann's equation \eqref{Bergmann} as an ordinary differential equation in the function space $\sF^1$. This requires some preparation.

Let $\xi:\R^2\to\C$ be an arbitrary function (tempered distribution), and $\tilde\xi(x,p)$ denote the partial Fourier transform of $\xi(x,y)$ with respect to $y$, i.e., $\tilde\xi(x,p):=\int_{-\infty}^\infty dy~e^{-ipy}\xi(x,y)$.
%	\be
%	\tilde\xi(x,p):=\int_{-\infty}^\infty dy~e^{-ipy}\xi(x,y).\nn
	%=\sqrt{2\pi}\,\br p|\psi(x)\kt,\nn
%	\ee
Then, for each $x\in\R$, the right-hand side of this relation defines a function of $p$ which we denote by $\tilde\xi(x)$. This is the element of $\sF^1$ that is given by  
	\be
	\big(\tilde\xi(x)\big)(p):=\tilde\xi(x,p).
	\label{t-xi}
	\ee
In particular, for every solution $\psi$ of the Bergmann equation \eqref{Bergmann}, $\tilde\psi(x)\in\sF^1$. 

Performing the partial Fourier transform of both sides of  \eqref{Bergmann} with respect to $y$, we find	 
	\begin{align}
    	&-\partial_x \big[\widehat{\alpha}(x)^{-1}\, \partial_x\tilde\psi(x)\big]
	+ \widehat \sV(x)\tilde \psi(x)=\widehat\varpi^2\tilde\psi(x),
	\label{t-Bergmann}
    	\end{align}
 where  
	\begin{align}
	&\widehat\alpha(x):=\alpha(x,\widehat y),
	&& \widehat\sV(x):= \widehat p\big[\widehat{\alpha}(x)^{-1}-\widehat I\big]\widehat p+
	k^2\big[\widehat I-\widehat{\beta}(x)\big],
	\label{sV=}\\
	&\widehat\beta(x):=\beta(x,\widehat y),
	&& \widehat\varpi:=\varpi(\widehat p),
	\label{hatvarpi-def}
	\end{align}
and $\widehat y$ and $\widehat p$ are respectively the $y$ components of the standard position and momentum operators of quantum mechanics in 2D. These operators act in $\sF^d$, with $d\in\{1,2\}$, according to 
	\begin{align}
	&(\widehat y\, \bF)(p)=i\partial_p\bF(p),
	&&(\widehat p\, \bF)(p)=p\,\bF(p).\nn
	\end{align} 
The second of these equations implies
	\be
	(\widehat\varpi\bF)(p)=\varpi(p)\bF(p).
	\label{hatvarpi=}
	\ee
Moreover, if $g:\R^2\to\C$ is a function, and $\tilde g(x,p)$ is the partial Fourier transforms of $g(x,y)$ with respect to $y$,  we can identify $g(x,\widehat y)$ with the following integral operator which acts in $\sF^d$.
	\be
	\big(g(x,\widehat y)\bF\big)(p):=\frac{1}{2\pi}\int_{-\infty}^\infty dq\:\tilde g(x,p-q)\bF(q).
	\label{integral-op}
	\ee
%Moreover, given a function $g:\R^2\to\C$,  we can identify $g(x,\widehat y)$ with the following integral operator that acts in $\sF^d$.
%	\be
%	\big(g(x,\widehat y)\bF\big)(p):=\frac{1}{2\pi}\int_{-\infty}^\infty dq\:\tilde g(x,p-q)\bF(q),
%	\label{integral-op}
%	\ee
%where $\tilde g(x,p)$ is the partial Fourier transforms of $g(x,y)$ with respect to $y$.
	
It is easy to determine asymptotic expressions for the solutions of \eqref{t-Bergmann} that are valid for $x\to\pm\infty$. To do this, first we use \eqref{condi-0} and \eqref{asym} to establish
	\be
	\tilde\psi(x,p)\to\frac{A_\pm(p) e^{i\varpi(p)x}+B_\pm(p) e^{-i\varpi(p)x}}{2\pi\varpi(p)}
	~~~~\for~~~~x\to\pm\infty.\nn
	%\label{asym2}
	\ee 
With the help of \eqref{t-xi} and \eqref{hatvarpi=}, we can express this relation in the form
	\be
	\tilde\psi(x)\to\frac{1}{2\pi}\,\widehat\varpi^{-1}\!\left(e^{ix\widehat\varpi}
	A_\pm +e^{-ix\widehat\varpi}B_\pm\right)
	~~~~\for~~~~x\to\pm\infty.
	\label{asym2}
	\ee  

Next, we introduce the following one-parameter family of elements of $\sF^2$.
	\be
	\Psi(x):=\pi\,e^{-ix\widehat\varpi_r \bsigma_3} 
	\left[\begin{array}{c} \widehat\varpi\,\tilde\psi(x)-i\widehat{\alpha}(x)^{-1}\,\partial_x\tilde\psi(x)\\
	\widehat\varpi\,\tilde\psi(x)+ i\widehat{\alpha}(x)^{-1}\,\partial_x\tilde\psi(x)\end{array}\right],
	\label{2-component}
	\ee
where $\bsigma_j$ with $j\in\{1,2,3\}$ denote the Pauli matrices, and
	\begin{align}
	&\widehat\varpi_r:=\widehat\Pi_k\widehat\varpi=\widehat\varpi\widehat\Pi_k,
	&& (\widehat\Pi_k\bF)(p):=\left\{\begin{array}{ccc}
	\bF(p)&\for&|p|<k,\\
	\bzero&\for&|p|\geq k.\end{array}\right.
	\label{project}
	\end{align}
If we use \eqref{asym2} and \eqref{2-component} to evaluate $\Psi(\pm\infty):=\lim_{x\to\pm\infty}\Psi(x)$, we find
	\be
	\Psi(\pm\infty)=\left[\begin{array}{c}
	A_\pm\\
	B_\pm\end{array}\right]\in\sF^2_k.
	\nn%\label{asym3}
	\ee
In light of this equation, we can express \eqref{M-def} as
	\be
	\widehat\bM\,\Psi(-\infty)=\Psi(+\infty).
	\label{M-Psi}
	\ee
	 	
The fact that $\tilde\psi(x)$ solves \eqref{t-Bergmann} is equivalent to the requirement that $\Psi(x)$ satisfies the time-dependent Schr\"odinger equation: 
	\be
    	i\partial_x \Psi(x) =\widehat \bH(x)\,\Psi(x),
	\label{sch-eq-t}
    	\ee
where $x$ plays the role of time, $\widehat \bH(x):\sF^2\to\sF^2$ is the effective Hamiltonian operator given by
	\begin{align}
	&\widehat\bH(x):=\frac{1}{2}e^{-ix\widehat\varpi_r\bsigma_3} 
	\left\{ \widehat\sV(x)\widehat\varpi^{-1}\bcK
	+\widehat\varpi[\widehat I-\widehat\alpha(x)]\bcK^T\right\}
	 e^{ix\widehat\varpi_r\bsigma_3}- i\widehat\varpi_i\bsigma_3,
	\label{bH=}\\
	&\bcK:=\left[\begin{array}{cc}
	1 & 1\\
	-1 & -1\end{array}\right],
	\quad\quad\quad\quad 
	\widehat\varpi_i:=-i(\widehat\varpi-\widehat\varpi_r)=
	-i\widehat\varpi(\widehat I-\widehat\Pi_k),
	\label{eta=}
	\end{align}
$\widehat I$ denotes the identity operators acting in $\sF^d$ %and $\sF^d_k$
 for $d\in\{1,2\}$, and the superscript $T$ stands for the transpose of a matrix. It is easy to see that, as a linear operator acting in the Hilbert space $L^2(\R)\otimes\C^{2\times 1}$, $\widehat\bH(x)$ is not Hermitian. Therefore, it generates a non-unitary dynamics in this Hilbert space. 

Let $\widehat\bU(x,x_0)$ denote the evolution operator associated with $\widehat\bH(x)$, i.e., the linear operator satisfying 
	\begin{align}
	&i\partial_x\widehat\bU(x,x_0)=\widehat\bH(x)
	\widehat\bU(x,x_0),
	&&\widehat\bU(x_0,x_0)=\widehat I.
	\nn
	\end{align}
Then, \eqref{M-Psi} implies $\widehat\bM=\widehat\bU(+\infty,-\infty)$. Moreover, repeating the analysis of Secs.~4 and 5 of Ref.~\cite{pra-2021}, we can show that
	\be
	\widehat\bM=\widehat\Pi_k\,\widehat\bcU(+\infty,-\infty)\widehat\Pi_k,
	\label{M=2}
	\ee
where $\widehat\bcU(x,x_0)$ is the evolution operator for the Hamiltonian operator:
	\be
	\widehat\bcH(x):=\frac{1}{2}e^{-ix\widehat\varpi\bsigma_3} 
	\left\{\widehat\sV(x)\widehat\varpi^{-1}\bcK
	+\widehat\varpi[\hat I-\widehat\alpha(x)]\bcK^T\right\}
	 e^{ix\widehat\varpi \bsigma_3}.
	\label{bcH=}
	\ee
	
For TE waves propagating in a nonmagnetic isotropic medium, $\alpha=\hat\mu=1$, 
$\beta=\hat\varepsilon$, $\widehat\alpha(x)=\widehat I$, $\widehat\sV(x)=k^2[\widehat I-\hat\varepsilon(x,\widehat y)]$,
%	\begin{align}
%	&\alpha=\hat\mu=1, 
%	&&\beta=\hat\varepsilon, 
%	&&\widehat\alpha(x)=\widehat I,
%	&&\widehat\sV(x)=k^2[\widehat I-\hat\varepsilon(x,\widehat y)],
%	\nn
%	\end{align}
and Eq.~\eqref{bcH=} reduces to Eq.~(7) of Ref.~\cite{pra-2025}.	

If there is some $\ell\in\R^+$ such that $\hat\varepsilon(x,y)=\hat\mu(x,y)=1$ for $x\notin[0,\ell]$, then for these values of $x$, $\widehat\sV(x)=\widehat 0$, and \eqref{bcH=} implies $\widehat\bcH(x)=\widehat 0$.\footnote{$\widehat 0$ stands for the zero operator acting in $\sF^d$ for $d\in\{1,2\}$.} Consequently, $\widehat\bcU(0,x)=\widehat I$ for $x<0$, and $\widehat\bcU(x,\ell)=\widehat I$ for $x>\ell$. In particular, $\widehat\bcU(+\infty,-\infty)=\widehat\bcU(\ell,0)$, and \eqref{M=2} becomes
	\be
	\widehat\bM=\widehat\Pi_k\,\widehat\bcU(\ell,0)\,\widehat\Pi_k.
	\label{M=new}
	\ee
Substituting the Dyson series for $\widehat\bcU(\ell,0)$ in this equation, we arrive at
	\begin{align}
	\widehat\bM&=\widehat\Pi_k\,
	\sT\exp\Big[-i\int_0^{\ell} dx\,\widehat\bcH(x)\Big]\widehat\Pi_k\nn\\
	&=\widehat\Pi_k+\sum_{n=1}^\infty(-i)^n\!\!
	\int_0^\ell \!dx_n\int_0^{x_n}\!\!dx_{n-1}\cdots\int_0^{x_2}\!\!dx_1\:\widehat\Pi_k
	\widehat\bcH(x_n)\widehat\bcH(x_{n-1})\cdots\widehat\bcH(x_1)\widehat\Pi_k,
	\label{dyson}
	\end{align}
where $\sT$ is the time-ordering operation with $x$ playing the role of time. Note that in view of \eqref{sV=}, \eqref{hatvarpi-def}, \eqref{integral-op}, \eqref{eta=}, \eqref{bcH=}, and \eqref{dyson}, $\widehat\sV(x)$, $\widehat\eta(x)$, $\widehat\bcH(x)$, and consequently the fundamental transfer matrix $\widehat\bM$ and its entries $\widehat M_{ab}$ are integral operators. 

This completes our outline of DFSS for the scattering problems defined by the Bergmann equation~\eqref{Bergmann} in 2D. It has the same features as the DFSS for potential scattering as described in Ref.~\cite{pra-2021}. The only difference is in the structure of the effective Hamiltonians \eqref{bH=} and \eqref{bcH=} which depend on $\alpha$ and $\beta$, or equivalently $\hat\varepsilon$ and $\hat\mu$. This analogy shows that the results pertaining exactness of the Born approximation in potential scattering \cite{pra-2021,jpa-2024} apply also for the scattering of waves fulfilling the Bergmann equation in 2D.

\section{Low-frequency scattering of TE and TM waves in 2D}
\label{S3}

Consider situations  where the inhomogeneities of the medium are confined to  the region given by $0\leq x\leq \ell$, as shown in Fig.~\ref{fig1}, and suppose that $\hat\varepsilon$ and $\hat\mu$ are functions of $x/\ell$ and $y$.\footnote{In general, $\hat\varepsilon$ and $\hat\mu$  also depend on $k$. We do not make this dependence explicit for brevity.} Then, there are functions $w_{\hat\varepsilon},w_{\hat\mu}:[0,1]\to\C$ such that 
	\begin{align}
	&\hat\varepsilon(x,y)=1+\left\{\begin{array}{ccc}
	w_{\hat\varepsilon}(\frac{x}{\ell},y)&\for&x\in[0,\ell],\\
	0&\for&x\notin[0,\ell],\end{array}\right.
	\label{epsilon-L}\\
	&\hat\mu(x,y)=1+\left\{\begin{array}{ccc}
	w_{\hat\mu}(\frac{x}{\ell},y)&\for&x\in[0,\ell].\\
	0&\for&x\notin[0,\ell].\end{array}\right.
	\label{mu-L}
	\end{align}
Substituting these formulas in \eqref{alpha-beta}, we find
	\begin{align}
	&\alpha(x,y)=1+\left\{\begin{array}{ccc}
	w_\alpha(\frac{x}{\ell},y)&\for&x\in[0,\ell],\\
	0&\for&x\notin[0,\ell],\end{array}\right.
	\label{alpha-L}\\
	&\beta(x,y)=1+\left\{\begin{array}{ccc}
	w_\beta(\frac{x}{\ell},y)&\for&x\in[0,\ell].\\
	0&\for&x\notin[0,\ell],\end{array}\right.
	\label{beta-L}
	\end{align}
where 
	\begin{align}
	&w_\alpha:=\left\{\begin{array}{ccc}
	w_{\hat\mu}&\for&\mbox{TE waves},\\
	w_{\hat\varepsilon}&\for&\mbox{TM waves},
	\end{array}\right.
	&&w_\beta:=\left\{\begin{array}{ccc}
	w_{\hat\varepsilon}&\for&\mbox{TE waves},\\
	w_{\hat\mu}&\for&\mbox{TM waves}.
	\end{array}\right.
	\label{w-alpha-beta}
	\end{align}

In the following, by ``low-frequency scattering'' we mean the scattering of incident waves whose wavenumber $k$ satisfies $k\ell\ll 1$. Notice that this is a weaker condition than the one adopted in the standard treatments of low-frequency scattering \cite{Stevenson-1953,Kriegsmann-1983,dassios-book,Ammari-2000} which demands the inhomogeneities of the medium to be confined to a disc (or sphere) of diameter $\ell$ such that $k\ell\ll 1$. Our aim is to construct the low-frequency series expansion of the scattering amplitude, i.e., 
	\be
	\ff(\theta)=\sum_{n=1}^\infty \ff^{(n)}(\theta)(k\ell)^n,
	\label{series}
	\ee
where $\ff^{(n)}$ are complex-valued functions that do not depend on $k\ell$.\footnote{The coefficients $\ff^{(n)}$ are functions of $\theta$, $\theta_0$, and $k$. For brevity we do not make their dependence on $\theta_0$ and $k$ explicit.} If we neglect all but the first $N$ terms of the right-hand side of \eqref{series}, we obtain the $N$-th order low-frequency approximation:
	\be
	\ff(\theta)\approx\sum_{n=1}^N \ff^{(n)}(\theta)(k\ell)^n.
	\label{LF-approx}
	\ee

We begin our discussion of the low-frequency scattering of TE and TM waves by introducing the following useful notation: For each $x\in\R$, we use $\cx$ to denote $x/\ell$,  i.e., $\check x:=x/\ell\in[0,1]$, and introduce the following dimensionless quantities.
	\begin{align}
	&\widehat{w}_\alpha(\cx):=w_\alpha(\cx,\widehat y)=
	\widehat\alpha(\ell\cx)-\widehat I,
	\qquad\qquad\qquad \quad~~~~~\:
	\widehat{w}_\beta(\cx):=w_\beta(\cx,\widehat y)=
	\widehat\beta(\ell\cx)-\widehat I,
	\label{c-alpha-beta}\\
	&v_\alpha(\cx,y):=1-\alpha(\ell\cx,y)^{-1}=\frac{w_\alpha(\cx,y)}{w_\alpha(\cx,y)+1}, 
	\quad\quad\qquad~ 
	\widehat v_\alpha(\cx):=v_\alpha(\cx,\widehat y)=
	\widehat I-\widehat\alpha(\ell\cx)^{-1},
	\label{v-def}\\
	&\widehat\sW(\check x):=-k^{-2}\widehat\sV(\ell\check x)= 
	k^{-2}\widehat p\: \widehat v_\alpha(\cx) \widehat p+
	\widehat{w}_\beta(\cx),
	\quad\quad\quad
	\widehat{\check\varpi}:=k^{-1}\widehat \varpi, 
	\label{W-def}\\
	&\widehat{\check\bcH}(\check x):=k^{-1}\widehat \bcH(\ell\check x)
	= -\frac{1}{2}e^{-ik\ell\,\check x\,\widehat{\check\varpi}\,\bsigma_3}
	\left[\widehat\sW(\check x)\bcK\,\widehat{\check\varpi}^{_{-1}}+ 
	\widehat{\check\varpi}\,\widehat{w}_\alpha(\cx)\,\bcK^T\right]
	e^{ik\ell\,\check x\,\widehat{\check\varpi}\,\bsigma_3}.
	\label{c-bcH-def}
	\end{align}
This allows us to express \eqref{dyson} in the form
	\begin{align}
	 \widehat\bM%&=\widehat\Pi_k \sT\exp\left\{-i\int_{0}^{1} d\check x\,k\ell\:\widehat{\check\bcH}(\check x) \right\}\widehat\Pi_k\nn\\
	 &=\widehat\Pi_k+\sum_{n=1}^\infty(-ik\ell)^n\!\!
	\int_0^1 \!d\cx_n\int_0^{\cx_n}\!\!d\cx_{n-1}\cdots\int_0^{\cx_2}\!\!d\cx_1\:\widehat\Pi_k
	\widehat{\check\bcH}(\cx_n)\widehat{\check\bcH}(\cx_{n-1})\cdots\widehat{\check\bcH}(\cx_1)\widehat\Pi_k.
	\label{dyson-L}
	 \end{align}
	 
It is easy to see from \eqref{c-bcH-def} that $\widehat{\check\bcH}(\check x)$ and consequently the integrand in \eqref{dyson-L} admit a power series expansion in nonnegative integer powers of $k\ell$. What is not so clear is that the structure of $\widehat{\check\bcH}(\check x)$ allows for a systematic calculation of the terms in these series. To see this, first we compute the products of the matrices $\bcK$, $\bcK^T$, $\bsigma_3$, and $\bsigma_\pm:=\bI\pm\bsigma_1$, and list them in Table~\ref{table1}. 
%	\begin{align}
%	&\bcK\bsigma_3=\bsigma_3\bcK^T=\bsigma_-, \quad\quad
%	\bsigma_3\bcK=\bcK^T\bsigma_3=\bsigma_+,\quad\quad
%	\bsigma_\pm^2=2\bsigma_\pm,
%	\label{id101}\\
%	&\bcK^2=\bcK\bsigma_-=\bsigma_+\bcK=\bsigma_-\bcK^T=\bcK^T\bsigma_+=\bsigma_\pm\bsigma_\mp=\bzero,
%	\label{id-102}\\
%	&\bcK\bsigma_+=\bsigma_-\bcK=2\bcK,\quad\quad
%	\bcK^T\bsigma_-=\bsigma_+\bcK^T=2\bcK^T,
%	\label{id103}\\
%	&\bcK\bcK^T=2\bsigma_-, \quad\quad
%	\bcK^T\bcK=2\bsigma_+,  \quad\quad
%	\bsigma_3\bsigma_+=\bcK,\quad\quad
%	\bsigma_3\bsigma_-=\bcK^T.
%	\label{id104}
%	\end{align}	
	\begin{table}
	\begin{center}
        \begin{tabular}{|c||c|c|c|c|c|} 
        \hline 
        $\cdot$ & $\bcK$ & $\bcK^T$ & $\bsigma_3$ & $\bsigma_+$ & $\bsigma_-$ \\ 
        \hline \hline
        $\bcK$ & $\bzero$ & $2\bsigma_-$ & $\bsigma_-$ & $2\bcK$ & $\bzero$  \\
        \hline
	$\bcK^T$ & $2\bsigma_+$ & $\bzero$ & $\bsigma_+$ & $\bzero$ &  $2\bcK^T$ \\
        \hline
        $\bsigma_3$ & $\bsigma_+$  & $\bsigma_-$ & $\bI$ & $\bcK$ & $\bcK^T$\\
        \hline
	$\bsigma_+$ & $\bzero$  & $2\bcK^T$ & $\bcK^T$ & $2\bsigma_+$ & $\bzero$\\
        \hline
        $\bsigma_-$ & $2\bcK$  & $\bzero$  & $\bcK$ & $\bzero$ & $2\bsigma_-$\\
        \hline
        \end{tabular}
        \caption{Multiplication table for the matrices $\bcK$, $\bcK^T$, $\bsigma_3$, and $\bsigma_\pm$. $\bzero$ and $\bI$ stand for the zero (null) and identity matrices, respectively.
        \label{table1}}
        \end{center}
     \end{table}
We have used the content of this table and Eq.~\eqref{c-bcH-def} to show that
	\be
	\widehat{\check\bcH}(\check x_n)\widehat{\check\bcH}(\check x_{n-1})	
	\cdots\widehat{\check\bcH}(\check x_1)=(-1)^n
	e^{-ik\ell\,\check x_n \widehat{\check\varpi}\,\bsigma_3}
	\widehat\bG_n\widehat\bG_{n-1}\cdots \widehat\bG_1,
	\label{Hn-1}
	\ee
where, for all $i\in\{1,2,\cdots n\}$,  
	\begin{align}
     	\widehat\bG_i&:=
	\frac{1}{2}\Big[\widehat\sW(\check x_i)\,\widehat{\check\varpi}^{_{-1}}\bcK+ 
	\widehat{\check\varpi}\,\widehat{w}_\alpha(\check x_i)\,
	\bcK^T\,\Big]
	e^{ik\ell\,\widehat{\check\varpi}(\check x_i-\check x_{i-1})\bsigma_3}\nn\\
    	%&=\frac{1}{2}\Big[
	%i\widehat\sW_i\,\widehat{\check\varpi}^{_{-1}} \widehat S_{i}\,\bsigma_-+
	%\widehat\sW_i\,\widehat C_{i}\widehat{\check\varpi}^{_{-1}}\bcK+
	%\widehat{\check\varpi} \,\widehat{w}_{\alpha\, i}\,\widehat C_{i}\,\bcK^T+
	%i\widehat{\check\varpi}\,\widehat{w}_{\alpha\, i}\,\widehat S_{i}\,\bsigma_+
	%\Big]\\
	&=\frac{1}{2}\Big[
	\widehat A_{i,11}\bsigma_-+
	\widehat A_{i,12}\widehat{\check\varpi}^{_{-1}}\bcK
	+ \widehat{\check\varpi} \,\widehat A_{i,21}\,\bcK^T
	+ \widehat{\check\varpi}\widehat A_{i,22}\,\widehat{\check\varpi}^{_{-1}} \bsigma_+
	\Big],
	\label{Gi-ded}
        	\end{align}
$\cx_0:=0$, and $\widehat A_{i,ab}$ are the entries of the operator $\widehat\bA_i:\sF^2\to\sF^2$ given by
	\begin{align}
	&\widehat\bA_i:=\left[\begin{array}{cc}
	i\widehat\sW_i\,\widehat{\check\varpi}^{_{-1}} \widehat S_{i} & 
	\widehat\sW_i\,\widehat C_{i}\\
	\widehat{w}_{\alpha\, i}\,\widehat C_{i} & 
	i\widehat{w}_{\alpha\, i}\,\widehat{\check\varpi}\,\widehat S_{i}
	\end{array}\right]=\widehat \bV(\cx_i)\widehat\bQ(\cx_i-\cx_{i-1}),
	\label{bA-def}\\
	&\widehat\sW_i:=\widehat\sW(\check x_i), \quad\quad\quad\quad\quad\quad\quad\;   
	\widehat{w}_{\alpha\, i}:=\widehat{w}_\alpha(\check x_i),
	\\
	&\widehat C_{i}:=\widehat C(\check x_i-\check x_{i-1}), \quad\quad\quad\quad\;\;  
	\widehat S_{i}:=\widehat S(\check x_i-\check x_{i-1}),  
    	\label{C-S-i} \\
	&\widehat C(\cx):=\cos(k\ell\,\cx\,\widehat{\check\varpi}), 
    	\quad\quad\quad\quad   \widehat S(\cx):=\sin(k\ell\,\cx\,\widehat{\check\varpi}),   
    	\label{CS-def}\\
	&\widehat\bV(\cx):=\left[\begin{array}{cc}
	\widehat\sW(\cx) & \widehat 0\\
	\widehat 0 & \widehat{w}_\alpha(\cx)\end{array}\right],\quad\quad
	\widehat\bQ(\cx):=\left[\begin{array}{cc}
	i\widehat{\check\varpi}^{\!_{-1}}\!\widehat S(\cx) & \widehat  C(\cx) \\
	\widehat  C(\cx) & i\widehat{\check\varpi} \widehat S(\cx)
	\end{array}\right].
	\label{VQ-def}
	\end{align}
	
A remarkable property of the operators $\widehat\bG_i$ is that 
	\be
	\widehat\bG_i\widehat\bG_{i-1}=\frac{1}{2}\Big[
	\widehat B_{i,11}\bsigma_-+
	\widehat B_{i,12}\widehat{\check\varpi}^{_{-1}}\bcK
	+ \widehat{\check\varpi} \,\widehat B_{i,21}\,\bcK^T
	+ \widehat{\check\varpi}\widehat B_{i,22}\,\widehat{\check\varpi}^{_{-1}} \bsigma_+
	\Big],
	\ee
where $\widehat B_{i,ab}$ are the entries of $\widehat\bB_i:=\widehat\bA_i\widehat\bA_{i-1}$. This shows that $\widehat\bG_n\widehat\bG_{n-1}\cdots \widehat\bG_1$ is given by the right-hand side of \eqref{Gi-ded} with $\widehat A_{i,ab}$ replaced with the entries of $\widehat \bA_n\widehat \bA_{n-1}\widehat \bA_1$. Using this observation, Eqs.~\eqref{Hn-1} and \eqref{bA-def}, and Table~\ref{table1}, we have shown that
	\begin{align}
	\widehat{\check\bcH}(\check x_n)\widehat{\check\bcH}(\check x_{n-1})	
	\cdots\widehat{\check\bcH}(\check x_1)=&\:\frac{(-1)^n}{2}\Big[
	\widehat L_{n,11}(\bx_n)\bsigma_-
	+\widehat L_{n,12}(\bx_n)\widehat{\check\varpi}^{_{-1}}\!\bcK\nn\\
	&\hspace{1.4cm}+\widehat{\check\varpi}\widehat L_{n,21}(\bx_n)\bcK^T
	+\widehat{\check\varpi}\widehat L_{n,22}(\bx_n)\widehat{\check\varpi}^{_{-1}}\! 
	\bsigma_+\Big],	
	\label{Hn-2n}
	\end{align}
where $\bx_n:=(\check x_{n},\cdots,\check x_{1})$,  $\widehat L_{n,ab}(\bx_n)$ are the entries of the operator $\widehat \bL_n(\bx_n):\sF^2\to\sF^2$ defined by
	\begin{align}
	&\widehat \bL_n(\bx_n):=\left\{\begin{aligned}
	&\widehat\bP(\cx_1)\widehat\bV(\cx_1)\widehat\bQ(\cx_1) &&\for~~n=1,\\
	&\widehat\bP(\cx_n)
	 \prod_{i=1}^{n}\widehat\bV(\cx_{i})\widehat\bQ(\cx_{i}-\cx_{i-1}) 
	&&\for~~n\geq 2,
	\end{aligned}\right.
	\label{bL-def}\\
	&\widehat\bP(\cx):=\left[\begin{array}{cc}
	\widehat C(\cx) & -i\widehat{\check\varpi}\widehat S(\cx)\\
	-i\widehat{\check\varpi}^{_{-1}}\! \widehat S(\cx) & \widehat C(\cx)\end{array}\right],
	\label{P-def}
	\end{align}
and $\prod_{i=1}^{n}\widehat\bV(\cx_{i})\widehat\bQ(\cx_{i}-\cx_{i-1})$ stands for the following ordered product of $\widehat\bV(\cx_{i})\widehat\bQ(\cx_{i}-\cx_{i-1})$'s.
	\[\widehat\bV(\cx_{n})\widehat\bQ(\cx_n-\cx_{n-1})
	\widehat\bV(\cx_{n-1})\widehat\bQ(\cx_{n-1}-\cx_{n-2})
	\cdots \widehat\bV(\cx_2)\widehat\bQ(\cx_2-\cx_1)\widehat\bV(\cx_1)\widehat\bQ(\cx_1).\]
%Note also that $\widehat\bQ(\cx)=-i\bsigma_2\widehat\bP(\cx)\widehat\bsigma_3$.
	
Substituting \eqref{Hn-2n} in \eqref{dyson-L}, we have
	\begin{align}
	\widehat\bM
	 &=\widehat\Pi_k+\frac{1}{2}\sum_{n=1}^\infty(ik\ell)^n 
	 \Big[	\widehat \sL_{n,11}\bsigma_-
	+\widehat \sL_{n,12}\widehat{\check\varpi}^{_{-1}}\!\bcK
	+\widehat{\check\varpi}\widehat\sL_{n,21}\bcK^T
	+\widehat{\check\varpi}\widehat\sL_{n,22}\widehat{\check\varpi}^{_{-1}}\! 
	\bsigma_+\Big],
	\label{dyson-L2}
	 \end{align}
where $\widehat\sL_{n,ab}$ are the entries of the operator $\widehat\bsL_n:\sF_k^2\to\sF_2^k$ defined by
	\be
	\widehat\bsL_n:=\int_0^1 \!d\cx_n\int_0^{\cx_n}\!\!d\cx_{n-1}\cdots\int_0^{\cx_2}\!\!d\cx_1\:\widehat\Pi_k \widehat\bL_n(\bx_n)\widehat\Pi_k.
	\label{bcL-def}
	 \ee
The fact that we can expand $\widehat C(\cx)$ and $\widehat S(\cx)$ in their McLaurin series,
	\begin{align}
	&\widehat C(\cx)=\sum_{m=0}^\infty\frac{(i k\ell\,\cx)^{2m}}{2m!}\,\widehat{\check\varpi}^{2m},
	&&i\widehat S(\cx)=\sum_{m=0}^\infty\frac{(ik\ell\,\cx)^{2m+1}}{(2m+1)!}\, \widehat{\check\varpi}^{2m+1},
	\label{CS-expand}
	\end{align}
together with Eqs.~\eqref{dyson-L}, \eqref{VQ-def}, and \eqref{Hn-2n} -- \eqref{bcL-def} show that $\widehat\bsL_n$, $\widehat{\check\bcH}(\check x_n)\widehat{\check\bcH}(\check x_{n-1})
	\cdots\widehat{\check\bcH}(\check x_1)$, and $\widehat\bM$ admit power series expansions in nonnegative integer powers of $k\ell$, i.e., low-frequency series expansions. 

%The following identities facilitate the task of identifying the coefficients of the even and odd terms in the series expansions of these operators in powers of $k\ell$,  i.e., their low-frequency series expansions.
%	\begin{align}
%	&\widehat\bP(\cx)=\widehat \bOmega\, \big[\widehat C(\cx)\bI-i\widehat S(\cx)\bsigma_3\big]\,\widehat \bOmega^{-1},
%	&&\widehat\bQ(\cx)=-i\bsigma_2\,\widehat\bP(\cx)\bsigma_3,
	%=\widehat \bOmega\, \big[\widehat C(\cx)\bI+\widehat S(\cx)\bsigma_3\big]\,\widehat \bOmega^{-1}\bsigma_3,
%	\\[6pt]
%	&\widehat C(\cx)=\sum_{m=0}^\infty\frac{(i k\ell\,\cx)^{2m}}{2m!}\,\widehat{\check\varpi}^{2m},
%	&&i\widehat S(\cx)=\sum_{m=0}^\infty\frac{(ik\ell\,\cx)^{2m+1}}{(2m+1)!}\, \widehat{\check\varpi}^{2m+1},
%	\label{CS-expand}
%	\end{align}
%where
%	\begin{align}
%	&\widehat\bOmega:=\left[\begin{array}{cc}
%	\widehat{\check\varpi} & \widehat{\check\varpi}\\
%	\widehat I & -\widehat I \end{array}\right],
%	&&\widehat\bOmega^{-1}=\frac{1}{2}\left[\begin{array}{cc}
%	\widehat{\check\varpi}^{\,_{-1}} & \widehat I \\
%	\widehat{\check\varpi}^{\,_{-1}} & -\widehat I \end{array}\right].
%	\end{align}
	
Let us denote the coefficients of the low-frequency series expansion of $\widehat\bsL_n$ by $\widehat\bsL_n^{\:(j)}$, so that 
	\be
	\widehat\bsL_n=\sum_{j=0}^\infty (k\ell)^j\widehat\bsL_n^{\:(j)}.
	\label{bcL-series}
	\ee
Substituting this equation in (\ref{dyson-L2}) and making use of \eqref{N-ab=}, we find
	\begin{align}
	&\widehat\bM=\widehat\Pi_k+\sum_{m=1}^\infty (k\ell)^m\, \widehat\bM^{(m)},
	&&\widehat N_{ab}=\sum_{m=1}^\infty (k\ell)^m \widehat N_{ab}^{(m)},
	\label{M-expand}
	\end{align}
where
	\begin{align}
	\widehat\bM^{(m)}&:=\frac{1}{2}\sum_{n=1}^m i^n
	 \Big[	\widehat \sL_{n,11}^{\:(m-n)}\bsigma_-
	+\widehat \sL_{n,12}^{\:(m-n)}\widehat{\check\varpi}^{_{-1}}\!\bcK
	+\widehat{\check\varpi}\widehat\sL_{n,21}^{\:(m-n)}\bcK^T
	+\widehat{\check\varpi}\widehat\sL_{n,22}^{\:(m-n)}\widehat{\check\varpi}^{_{-1}}\! 
	\bsigma_+\Big],
	\label{dyson-L3}\\
	\widehat N_{ab}^{(m)}&:=\frac{1}{2}\sum_{n=1}^m i^n
	 \Big[	(-1)^{a+b-1}\widehat \sL_{n,11}^{\:(m-n)}
	 +(-1)^a\widehat \sL_{n,12}^{\:(m-n)}\widehat{\check\varpi}^{_{-1}}
	+(-1)^b\widehat{\check\varpi}\widehat\sL_{n,21}^{\:(m-n)}
	-\widehat{\check\varpi}\widehat\sL_{n,22}^{\:(m-n)}\widehat{\check\varpi}^{_{-1}}\Big].
	\label{Nab-m=2}
	 \end{align}	
	  
In Appendix~A, we obtain explicit formulas for $\widehat N_{ab}^{(1)}$ and $\widehat N_{ab}^{(2)}$, and use them together with \eqref{f=} and \eqref{B-L-series} -- \eqref{A-R-series} to determine the leading- and next-to-leading-order terms in the low-frequency expansion of the scattering amplitude \eqref{series}. The result is 
	\begin{align}
	\ff^{(1)}(\theta)=&\frac{k}{2\sqrt{2\pi}}
	\Big[W_0(k\,\fs,k\,\fs_0)+\fc_0\fc \,
	\overline{\tilde w}_{\alpha,0}\big(k(\fs-\fs_0)\big)\Big],
	\label{f1=}\\
	\ff^{(2)}(\theta)=&\frac{ik}{2\sqrt{2\pi}}\Bigg\{
	\fc_0\Big[X_1(k\fs,k\fs_0)+X_2\big(k(\fs-\fs_0)\big)+W_1(k\fs,k\fs_0)
	-\fc^2\overline{\tilde w}_{\alpha,1}\big(k(\fs-\fs_0)\big)\Big]\nn\\
	&\hspace{1.8cm}+\fc\Big[Y_1(k\fs,k\fs_0)+Y_2\big(k(\fs-\fs_0)\big)-W_1(k\fs,k\fs_0)
	+\fc_0^2\overline{\tilde w}_{\alpha,1}\big(k(\fs-\fs_0)\big)\Big]\nn\\
	&\hspace{1.8cm}+\frac{k}{4\pi}\int_{-\frac{\pi}{2}}^{\frac{\pi}{2}}d\varphi
	\Big[W_0(k\fs,k\sin\varphi)-\fc\cos\varphi\,\overline{\tilde w}_{\alpha,0}
	\big(k(\fs-\sin\varphi)\big)\Big]
	\times\nn\\
	&\hspace{4.2cm}\Big[W_0(k\sin\varphi,k\fs_0)
	-\fc_0\cos\varphi\,\overline{\tilde w}_{\alpha,0}\big(k(\sin\varphi-\fs_0)\big)\Big]\Bigg\},
	\label{f2=}
	\end{align}
where
	\begin{align}
	&\fc:=\cos\theta,\qquad
	\fs:=\sin\theta, \qquad
	\fc_0:=\cos\theta_0, \qquad 
	\fs_0:=\sin\theta_0,
	\label{ccss-def}\\
	& W_l(p,p'):=k^{-2}p\,p'\:\overline{\tilde v}_{\alpha,l}(p-p')+\overline{\tilde w}_{\beta,l}(p-p'),
	\label{W-def}\\	
	&\overline{\tilde w}_{\alpha,l}(p):=\int_0^1d\cx\:\cx^l\:\tilde w_{\alpha}(\cx,p),
	\qquad\qquad
	\overline{\tilde w}_{\beta,l}(p):=\int_0^1d\cx\:\cx^l\:\tilde w_{\beta}(\cx,p),
	\label{w-ab-def}\\
	&\overline{\tilde v}_{\alpha,l}(p):=\int_0^1d\cx\:\cx^l\:\tilde v_{\alpha}(\cx,p),
	\label{v=}\\
		& {X}_1(p,p'):=\frac{p}{2\pi\,k^2}\int_0^1\!\! d\cx_2\int_0^{\cx_2}\!\!\! d\cx_1
	\int_{-\infty}^\infty \!\!\! dq~q\,\tilde v_\alpha(\cx_2,p-q)\tilde w_\alpha(\cx_1,q-p'),\\
	& {X}_2(p):=\frac{1}{2\pi}\int_0^1\!\!d\cx_2\int_0^{\cx_2}\!\!\! d\cx_1
	\int_{-\infty}^\infty \!\!\! dq~ \tilde w_\beta(\cx_2,p-q)\tilde w_\alpha(\cx_1,q),\\
	& {Y}_1(p,p'):=\frac{p'}{2\pi\,k^2}\int_0^1\!\! d\cx_2\int_0^{\cx_2}\!\!\!d\cx_1
	\int_{-\infty}^\infty\!\!\! dq~q\,\tilde w_\alpha(\cx_2,p-q)\tilde v_\alpha(\cx_1,q-p'),\\
	& {Y}_2(p):=\frac{1}{2\pi}\int_0^1\!\! d\cx_2\int_0^{\cx_2}\!\!\! d\cx_1
	\int_{-\infty}^\infty \!\!\! dq~ \tilde w_\alpha(\cx_2,p-q)\tilde w_\beta(\cx_1,q).
	\label{Y2-def}
	\end{align}
Observe that according to \eqref{W-def},
	\be
	W_l(k\fs,k\fs_0)=\fs_0 \fs\,\overline{\tilde v}_{\alpha,l}\big(k(\fs-\fs_0)\big)+
	\overline{\tilde w}_{\beta,0}\big(k(\fs-\fs_0)\big).
	\label{WL-29}
	\ee

\section{Application to a class of exactly solvable gratings}
\label{S4}

Consider the medium $\sM$ defined by 
	\begin{align}
	&\hat\varepsilon(x,y)=1+\left\{\begin{array}{ccc}
	\fz_0(\frac{x}{\ell})+\fz_1(\frac{x}{\ell})\, e^{i\fK\, y} &\for&x\in[0,\ell],\\[3pt]
	0&\for&x\notin[0,\ell],\end{array}\right.
	&&\hat\mu(x,y)=1,
	\label{epsilon-exact}
	\end{align}	 
where $\fz_0$ and $\fz_1$ are possibly complex-valued functions of $x/\ell$ (and possibly $k$), and $\fK$ is a positive real parameter. In Ref.~\cite{pra-2017}, we use the dynamical formulation of potential scattering in two dimensions to obtain the exact solution of the scattering problem for the normally-incident TE waves scattered by $\sM$. In particular, we show that the scattering amplitude has the form
	\be
	\ff(\theta)=\sum_{j=0}^{\lfloor k/\fK \rfloor} \big[\tau_{j+}\delta(\theta-\theta_{j+})+
	\tau_{j-}\delta(\theta-\theta_{j-})\big],
	\ee
where $\lfloor k/\fK \rfloor$ stands for the integer part of $k/\fK$, $\tau_{j\pm}$ are complex amplitudes admiting analytic expressions, and $\theta_{j\pm}$ are the angles given by
	\begin{align}
	%&\theta_{j+}:=\arcsin\big(\tfrac{j\fK}{k}\big)\in[0,\tfrac{\pi}{2}),
	%&\theta_{j-}:=\pi-\theta_{j+}\in(\tfrac{\pi}{2},\pi].
	&\theta_{j+}:=\arcsin\big(\tfrac{j\fK}{k}\big)\in[0^\circ,90^\circ),
	&\theta_{j-}:=180^\circ-\theta_{j+}\in(90^\circ,180^\circ].
	\end{align}
	
For the special cases, where $\fz_0$ and $\fz_1$ are constant, and the imaginary part of $\fz_0$ is not smaller than $|\fz_1|$, $\sM$ corresponds to the lossy diffraction grating studied by Berry in Ref.~\cite{berry-1998}.%\footnote{Berry's results hold for TE waves with arbitrary incidence angle.}

In the following we explore the utility of the results of the preceding section in the description of the scattering of low-frequency TM waves by $\sM$ for cases where $\fz_0$ and $\fz_1$ are possibly $k$-dependent constant coefficients. 
%and the angle of incidence is chosen to coincide with Brewster's angle. 
In view of \eqref{w-alpha-beta} and \eqref{epsilon-exact}, this  corresponds to setting
	\begin{align}
	&w_{\hat\varepsilon}(\cx,y):=\fz_0+\fz_1 e^{i\fK\,y},
	&&w_{\hat\mu}(\cx,y):=0,
	\label{ws-exact}
	\end{align}
in \eqref{epsilon-L} and \eqref{mu-L}.  %This choice of incidence angle allows us to delineate the effects of the complex periodic character of the permittivity profile. It also simplifies the calculations of the scattering amplitude. 

The general scattering problem for TE and TM waves scattered by a medium given by \eqref{epsilon-exact} belongs to a larger class of exactly solvable scattering problems whose treatment is beyond the scope of the present article. In Supplementary Materials \cite{SM}, we provide a brief summary of the steps of the exact evaluation of the scattering amplitude for the scattering of TM waves of wavenumber $k\leq\fK$ at Brewster's angle of incidence by the medium given by \eqref{epsilon-L}, \eqref{mu-L}, and \eqref{ws-exact} with $\fz_0$ real and positive, and $|\fz_1|<|\fz_0+1|$.{\footnote{{This condition holds for non-exotic material.}}} This allows us to provide a graphical comparison of the effectiveness of the first- and second-order low-frequency approximations for this model. 

For the scattering of TM waves by a nonmagnetic scatterer, $\alpha=\hat\varepsilon$ and $\beta=\hat\mu=1$. These equations together with \eqref{v-def}, \eqref{W-def} -- \eqref{Y2-def}, \eqref{ws-exact}, and $|\fz_1|<|\fz_0+1|$ imply 
	\begin{align}
	&{\tilde w}_{\alpha}(p)=\overline{\tilde w}_{\alpha,0}(p)
	=2\,\overline{\tilde w}_{\alpha,1}(p)=
	2\pi\big[\fz_0\delta(p)+\fz_1\delta(p-\fK)\big],
	\label{eq-z1}\\
	&{\tilde v}_{\alpha}(p)=\overline{\tilde v}_{\alpha,0}(p)=2\,\overline{\tilde v}_{\alpha,1}(p)=2\pi
	\sum_{j=0}^{\lfloor p/\fK \rfloor}\fa_j\,\delta(p-j\fK),\\
	&\overline{\tilde w}_{\beta,l}(p)=X_2(p)=Y_2(p)=0,\\
	%&\begin{aligned}
	%W_0(p,p')&=2W_1(p,p')=k^{-2}p\,p' \tilde v_\alpha(p-p')\\
	%&=\frac{2\pi}{k^2}\sum_{j=0}^{\lfloor(p-p')/\fK \rfloor}\!\!\! \fa_j\, p(p-j\fK)\delta(p-p'-j\fK),
	%\end{aligned}\\
	&W_0(p,p')=2W_1(p,p')=\frac{p\,p'\tilde v_\alpha(p-p')}{k^2}=
	\frac{2\pi p'}{k^2}\sum_{j=0}^{\lfloor(p-p')/\fK \rfloor}\!\!\! \fa_j\,(p'+j\fK) \delta(p-p'-j\fK),\\	
	&\begin{aligned}
	X_1(p,p')
	%&=\tfrac{p}{2k^2}\big[\fz_0 p'\,\tilde v_\alpha(p-p')+\fz_1(p'+\fK)\tilde v_\alpha(p-p'-\fK)\big]\\
	&=\tfrac{1}{2}\big[\fz_0 W_0(p,p')+\fz_1 W_0(p,p'+\fK)\big],
	\end{aligned}\\[6pt]
	&\begin{aligned}
	Y_1(p,p')
	%&=\tfrac{p'}{2k^2}\big[\fz_0 p\,\tilde v_\alpha(p-p')+\fz_1(p-\fK)\tilde v_\alpha(p-p'-\fK)\big]\\
	&=\tfrac{1}{2}\big[\fz_0 W_0(p,p')+\fz_1 W_0(p-\fK,p')\big],
	\end{aligned}
	\label{eq-z2}
	\end{align}
where 
	\begin{align}
	&\fa_0:=\frac{\fz_0}{\fz_0+1}, &&
	\fa_j:=\frac{(-1)^{j+1}\fz_1^j}{(\fz_0+1)^{j+1}}~~\for~~j\geq 1.
	\label{an-def}
	\end{align} 
	
Substituting \eqref{eq-z1} -- \eqref{eq-z2} in \eqref{f1=} and \eqref{f2=}, and making use of \eqref{ccss-def} and the identity
	\[\delta(\sin\theta-\sin\varphi)=\frac{\delta(\theta-\varphi)+\delta(\theta+\varphi-\pi)}{|\cos\varphi|}~~\for~~\theta,\varphi\in(-\tfrac{\pi}{2},\tfrac{\pi}{2})\cup(\tfrac{\pi}{2},\tfrac{3\pi}{2}),\]
we find
	\begin{align}
	&\ff^{(n)}(\theta)=\sum_{j=0}^{J} \Big[\tau^{(n)}_{j+}\delta(\theta-\theta_{j+})+
	\tau^{(n)}_{j-}\delta(\theta-\theta_{j-})\Big],
	\label{f=shift}
	\end{align}
where $n\in\{1,2\}$, and 
	\begin{align}
	%&\theta_{j+}:=\arcsin\fs_j\in(-\tfrac{\pi}{2},\tfrac{\pi}{2}),%\qquad\qquad
	%\qquad \theta_{j-}:=\pi-\theta_{j+},%\in(\tfrac{\pi}{2},\tfrac{3\pi}{2}),
	&J:=\lfloor \tfrac{k}{\fK}(1-\sin\theta_0)\rfloor,\quad\qquad
	\theta_{j+}:=\arcsin\fs_j\in(-90^\circ,90^\circ),
	\label{theta-j}\\	
	&\fs_j:=\sin\theta_0+\frac{j\fK}{k},\qquad\qquad~
	\theta_{j-}:=180^\circ-\theta_{j+},%\in(\tfrac{\pi}{2},\tfrac{3\pi}{2}),
	\label{sj-def} 	
	\\[3pt]
	&\tau^{(1)}_{j\pm}:=\sqrt{\frac{\pi}{2}}\times
	\left\{\begin{array}{cc}
	\fa_0\fs_0^2\sec\theta_{0+} \pm \fz_0\,\fc_0&\for~~j=0,\\
	\fa_1\fs_0\fs_1\sec\theta_{1+}\pm\fz_1\fc_0&\for~~j=1,\\
	\fa_j\fs_0\fs_j\sec\theta_{j+} &\for~~j\geq 2,
	\end{array}\right.
	\label{tau1-j-pm}\\
	%\end{align}\vspace{-12pt}
	&\begin{aligned}
	\tau^{(2)}_{0\pm}:=&\frac{i}{2}\sqrt{\frac{\pi}{2}}
	\Big\{\sec\theta_{0+}\big[\fa_0\fs_0^2(\fa_0\fs_0^2\sec\theta_{0+}+\fc_0)-\fz_0\fc_0^3\big]
	\\
	&\hspace{1.5cm}
	\pm\big[\fz_0\fc_0(\fz_0\cos\theta_{0+}+\fc_0)-\fa_0\fs_0^2\big]\Big\},
	\end{aligned}
	\label{tau2-zero-pm}\\
	&\begin{aligned}
	\tau^{(2)}_{1\pm}:=&\frac{i}{2}\sqrt{\frac{\pi}{2}}
	\Big\{\fa_1\fs_0\fs_1\sec\theta_{1+}\big[\fa_0(\fs_0^2\sec\theta_{0+}
	+\fs_1^2\sec\theta_{1+})+\fc_0\big]-\fz_1\fc_0\cos\theta_{1+} \\
	&\hspace{1.5cm}
	\pm\Big(\fz_1\fc_0\big[\fz_0(\cos\theta_{0+}+\cos\theta_{1+})+\fc_0)\big]
	-\fa_1\fs_0\fs_1\Big)\Big\}.
	\end{aligned}
	\label{tau2-1-pm}
	\end{align}
Analytic expressions for $\tau^{(2)}_{j\pm}$ with $j\geq 2$ can also be obtained in a similar manner. We do not include them here, because they are too lengthy to be of immediate use. More importantly, because we confine our attention to incident waves with $k\leq\fK$, they do not contribute to the outcome of the first- and second-order low-frequency approximations. 
		 	 
It is important to realize that according to \eqref{theta-j},
	\be
	\sin\theta_0<1-\frac{J\fK}{k}.
	\label{condi-k}
	\ee 
For $k\leq\tfrac{\fK}{2}$, this implies $\sin\theta_0< 1-2J$, which can only be met for $J=0$. This means that the system does not deflect the incident waves with wavenumbers not exceeding $\tfrac{\fK}{2}$. For $\tfrac{\fK}{2}<k\leq\fK$, \eqref{condi-k} gives $\sin\theta_0< 1-J$ which implies $J\in\{0,1\}$. Furthermore, for $J=1$, \eqref{condi-k} holds only if 
	\be
	-90^\circ<\theta_0<-\theta_{0\star}~~~{\rm or}~~~180^\circ+\theta_{0\star}<\theta_0<270^\circ,
	\label{condi-94}
	\ee
where 
	\be
	\theta_{0\star}:=\arcsin\left(\tfrac{\fK}{k}-1\right). 
	\label{theta0-star}
	\ee

\begin{figure}
        \begin{center}
        \includegraphics[scale=.3]{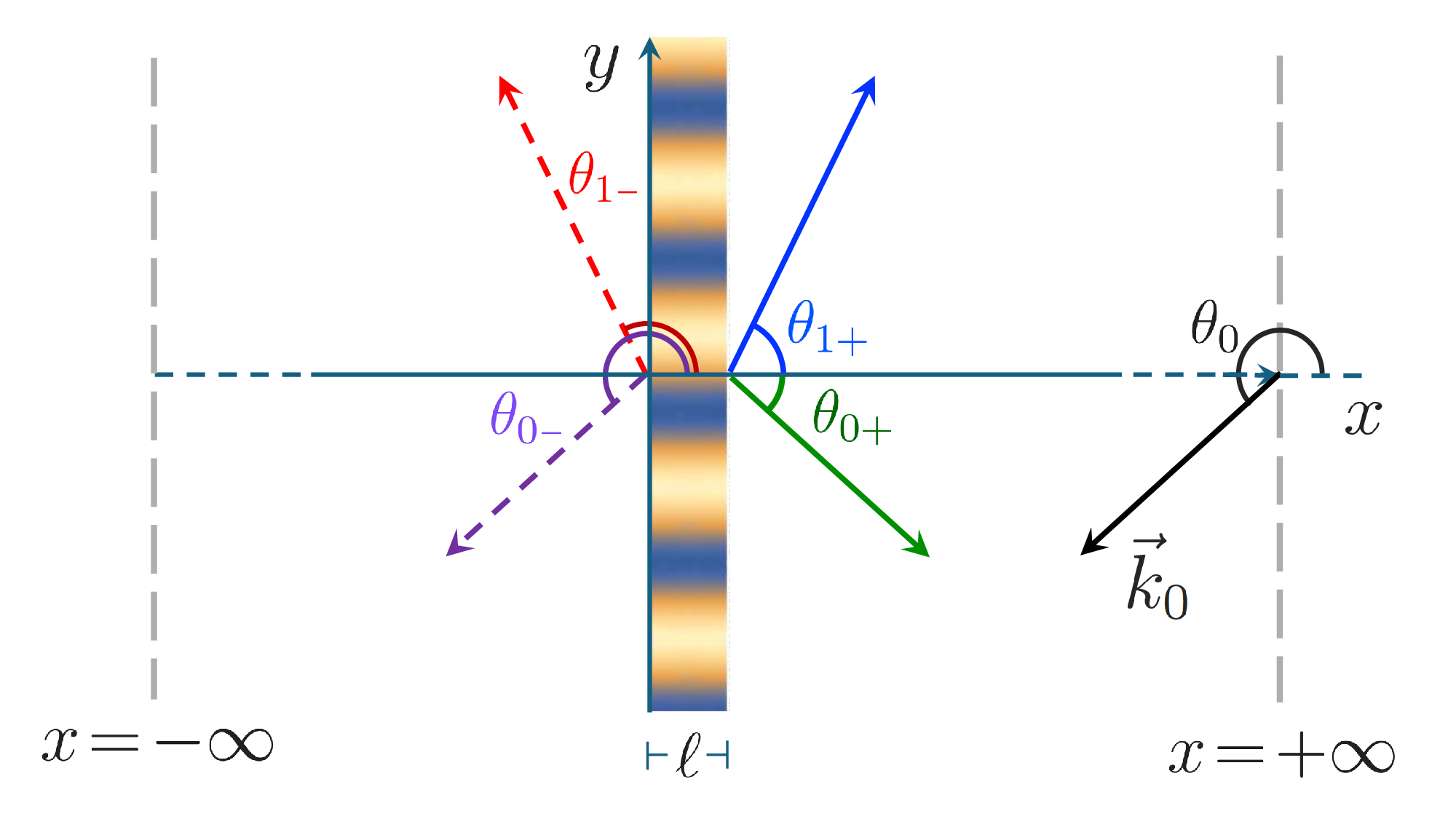} 
        \caption{{Schematic views of the scattering of a TM wave with wavenumber $k\in(\kappa_0,\fK]$ by a grating given by \eqref{ws-exact}. A density plot of the modulus of the relative permittivity of the grating, i.e., $|\hat\varepsilon(x,y)|$, is used to color the region representing the grating. The arrows mark the incident and scattered wave vectors. $\theta_{0\pm}$ and $\theta_{1\pm}$ are the angles the scattered wave vectors make with the positive $x$ axis.}}
        \label{fig2}
        \end{center}
        \end{figure}

Next, consider the homogeneous slab obtained by taking $\fz_0$ to be a positive real constant and $\fz_1=0$ in \eqref{epsilon-exact}. Let $\theta_B$ stand for Brewster's angle for this slab \cite{born-wolf}, i.e.,
	\be
	\theta_B:=\arctan\sqrt{\fz_0+1}.
	\label{Brewster}
	\ee
Then the slab does not reflect TM waves with incidence angles 
$\pm\theta_B$ and $180^\circ\pm\theta_B$. This observation suggests that these choices for the incidence angle facilitate the identification of the scattering contributions associated with the $y$-dependent complex periodic part of the permittivity profile \eqref{epsilon-exact}. Moreover, in view of  \eqref{condi-94}, we can ensure the presence of the deflected waves corresponding to the terms with $j=1$ in \eqref{f=shift} by taking
	\begin{align}
	&\theta_0=180^\circ+\theta_B
	\label{incidence-angle}
	\end{align}
and demanding that $\theta_{0\star}<\theta_B$ and $k\leq\fK$. Using \eqref{theta0-star} and \eqref{Brewster}, we can express these conditions in the form
	\be
	\kappa_0<k\leq \fK,
	\label{condi-123}
	\ee
where
	\be
	\kappa_0:=\frac{\fK}{1+\sqrt{\frac{\fz_0+1}{\fz_0+2}}}.
	\label{kappa-zero}
	\ee	
Because $\fz_0\in\R^+$, $\kappa_0>\fK/2$. This shows that if the incidence angle and wavenumber respectively satisfy \eqref{incidence-angle} and  \eqref{condi-123}, the deflected waves are generically present. Under these conditions, the scattering amplitude for the low-frequency TM waves is given by \eqref{f=shift} with $J=1$,  $\tau^{(n)}_{j+}$ and $\tau^{(n)}_{j-}$ respectively giving the amplitudes of the reflected and transmitted waves, and
	\begin{align}
	&\theta_{0+}=-\theta_B=-\arctan\sqrt{\fz_0+1}, 
	&& \theta_{0-}=180^\circ+\theta_B=\theta_0,
	\label{thetas-zero}\\
	&\theta_{1+}=\arcsin\left(\tfrac{\fK}{k}-\sin\theta_B\right),
	&&\theta_{1-}=180^\circ-\theta_{1+}.
	\label{thetas-1}
	%=\arcsin\left[\frac{\fK}{k}-\frac{\fz_0+1}{\sqrt{(\fz_0+1)^2+1}}\right],\nn
	\end{align} 
Fig.~\ref{fig2} illustrates the incident and scattered wave vectors and the angles $\theta_0$, $\theta_{0\pm}$, and $\theta_{1\pm}$.
	
\begin{figure}
        \begin{center}
        \includegraphics[scale=.65]{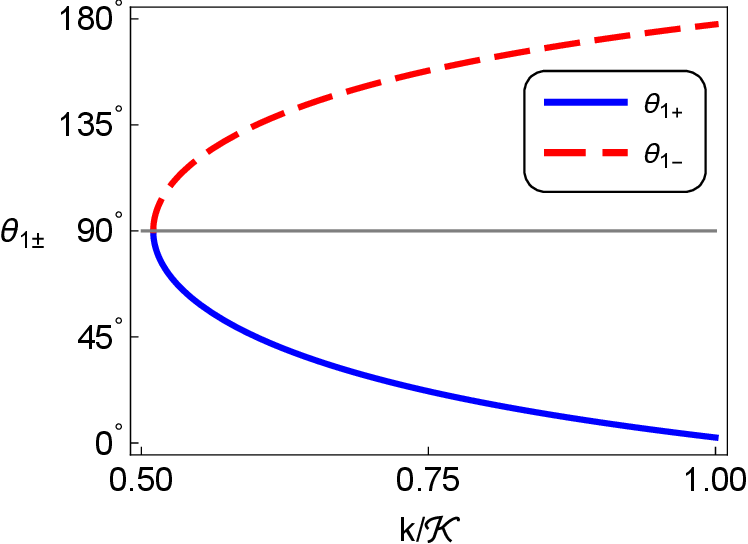} 
        \caption{Graph of $\theta_{1\pm}$ as a function of $k/\fK$ for $\fz_1$ and $\fz_2$ given by \eqref{specs}. The two curves meet at $k/\fK=\kappa_0/\fK=0.510$.}
        \label{fig3}
        \end{center}
        \end{figure}

In view of \eqref{ccss-def}, \eqref{Brewster}, \eqref{incidence-angle}, and \eqref{thetas-zero}, 
	\begin{align}
	&\sin\theta_{0+}=-\fs_0=\sin\theta_B=\sqrt{\frac{\fz_0+1}{\fz_0+2}},
	&&\fc_0=-\cos\theta_B=-\frac{1}{\sqrt{\fz_0+2}}.
	\label{sins-cos}
	\end{align}
Substituting these formulas in \eqref{tau1-j-pm} and \eqref{tau2-zero-pm}, and using \eqref{an-def}, we find
	\begin{align}
	&\tau_{0+}^{(1)}=\tau_{0+}^{(2)}=0, 
	&&\tau_{0-}^{(1)}=\frac{\sqrt{2\pi}\,\fz_0}{\sqrt{\fz_0+2}},
	&&\tau_{0-}^{(2)}=i\sqrt{\frac{\pi}{2}}\left(\frac{\fz_0^2}{\fz_0+2}\right).
	\label{tau-zero=}
	\end{align}
The first of these relations is consistent with the fact that for $\fz_1=0$ the slab does not reflect the wave. We can also express $\tau_{1\pm}^{(1)}$ and $\tau_{1\pm}^{(2)}$ in terms of $\fz_0$ and $\fz_1$, but these also involve $\theta_{1+}$ which is a function of $k/\fK$.  In Fig.~\ref{fig3}, we plot $\theta_{1+}$ as a function of $k/\fK$ for 
	\begin{align}
	&\fz_0=10.58,
	&&\fz_1=0.07,
	\label{specs}
	\end{align} 
which correspond to an InGaAsP slab whose refractive index is given by $\fn=3.4+i\kappa$, $\kappa\in\R$, and $|\kappa|\leq 0.01$. Substituting \eqref{specs} in  \eqref{Brewster}, \eqref{kappa-zero}, \eqref{thetas-zero}, and \eqref{tau-zero=}, we find 
	\begin{align}
	&\kappa_0=0.510\fK,
	&&\theta_{0+}=-\theta_B=-73.6^\circ, && 
	\theta_{0-}=\theta_0=253.6^\circ, \\
	&\tau_{0-}^{(1)}=7.48, &&\tau_{0-}^{(2)}=11.15 i.
	\label{tau0m=}
	\end{align}

According to \eqref{series} and \eqref{f=shift}, for $k\leq\fK$, 
	\begin{align}
	\ff(\theta)=\sum_{j=0}^1 \big[\tau_{j+}\delta(\theta-\theta_{j+})+
	\tau_{j-}\delta(\theta-\theta_{j-})\big],
	\end{align}
where
	\be
	\tau_{j\pm}:=\sum_{n=1}^\infty \tau_{j\pm}^{(n)}(k\ell)^n.
	\label{tau-exact}
	\ee
The $N$-th order low-frequency approximation  \eqref{LF-approx} corresponds to 
	\be
	\tau_{j\pm}\approx\sum_{n=1}^N \tau_{j\pm}^{(n)}(k\ell)^n.
	\label{tau-approx}
	\ee
According to \eqref{tau-zero=}, \eqref{tau0m=}, and \eqref{tau-approx}, for $j=0$, the first- and second-order low-frequency approximations give 
	\begin{align}
	&\tau_{0+}\approx 0,
	&&\tau_{0-}\approx \left\{\begin{array}{ccc}
	7.48\,k\ell&\for&1^{\rm st}\,\mbox{order approximation},\\
	7.48\,k\ell+11.15 i(k\ell)^2&\for&2^{\rm nd}\,\mbox{order approximation}.\end{array}\right.
	\label{tau-zero-approx}
	\end{align}
	
\begin{figure}
        \begin{center}
        \includegraphics[scale=.3]{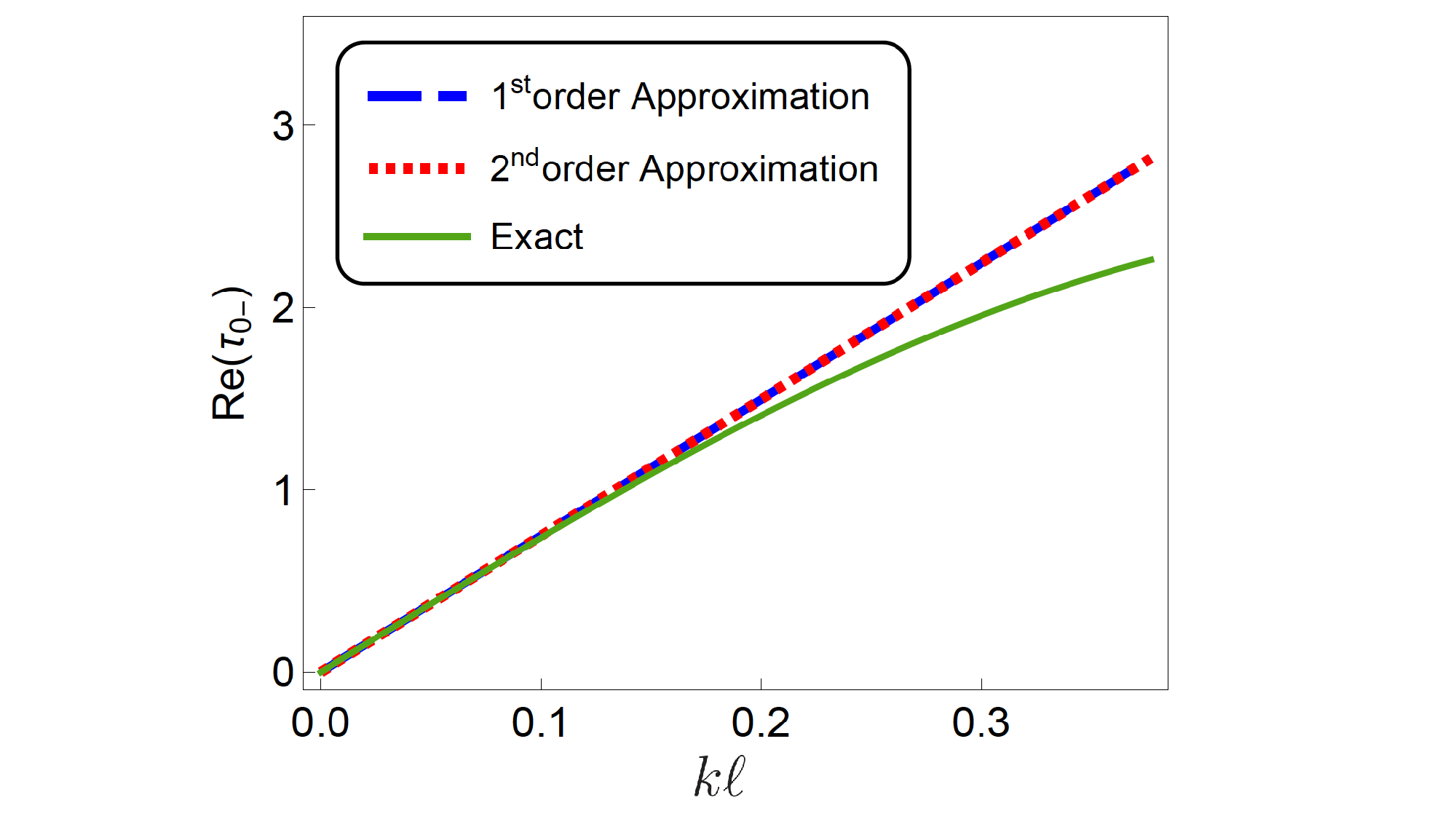}~\hspace{-2cm}~\includegraphics[scale=.3]{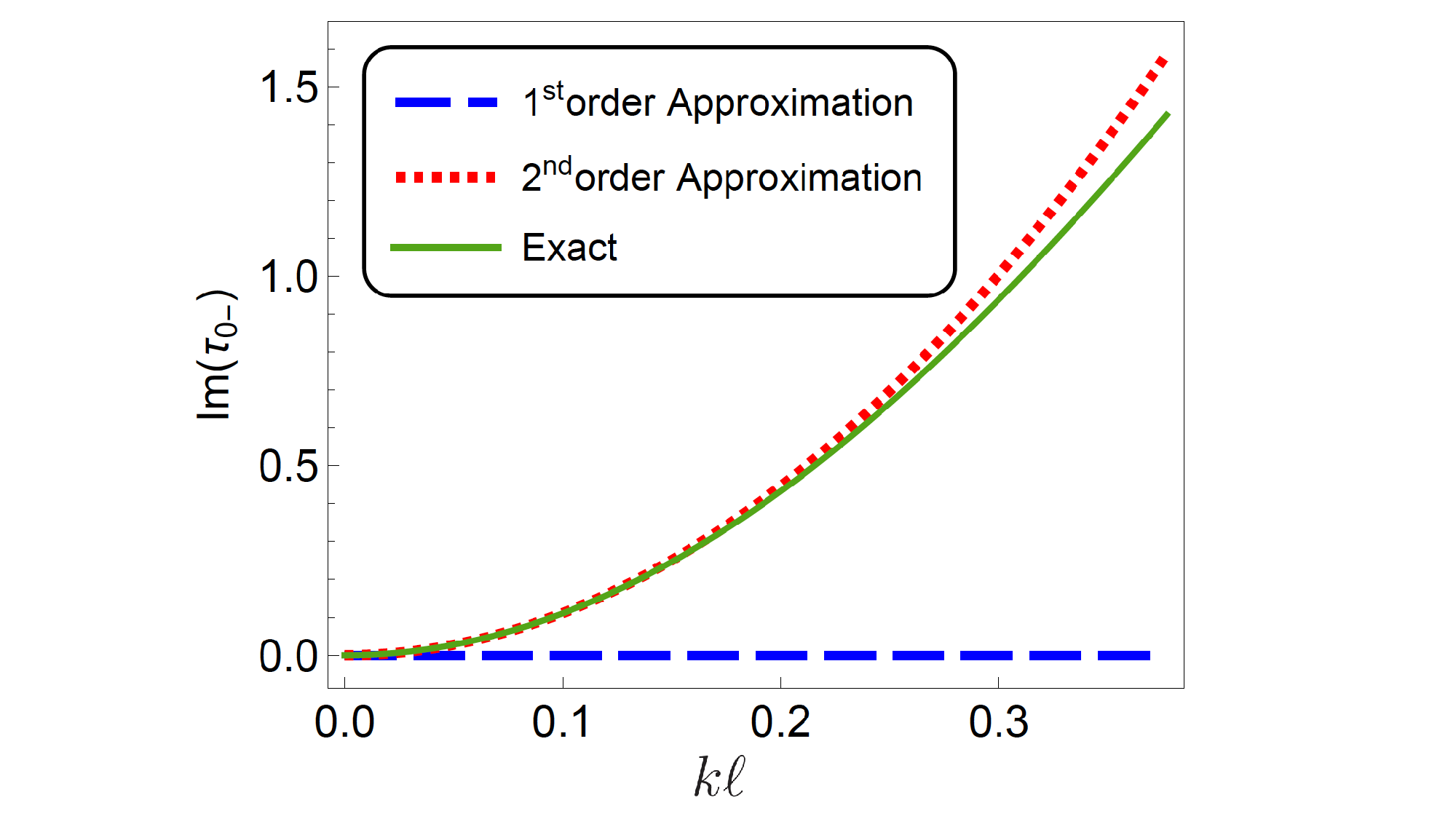}
       \caption{Graphs of the real and imaginary parts of $\tau_{0-}$ as a function of $k\ell$ for $\fz_1$ and $\fz_2$ given by \eqref{specs}, $\ell=100~{\rm nm}$, and $\fK=\pi~\mu{\rm m}^{-1}$. The dashed and dotted curves respectively correspond to the outcome of the first- and second-order low-frequency approximations, while the solid curve corresponds to the exact calculation of $\tau_{0-}$.}
        \label{fig4}
        \end{center}
        \end{figure}

\begin{figure}
        \begin{center}
        \includegraphics[scale=.3]{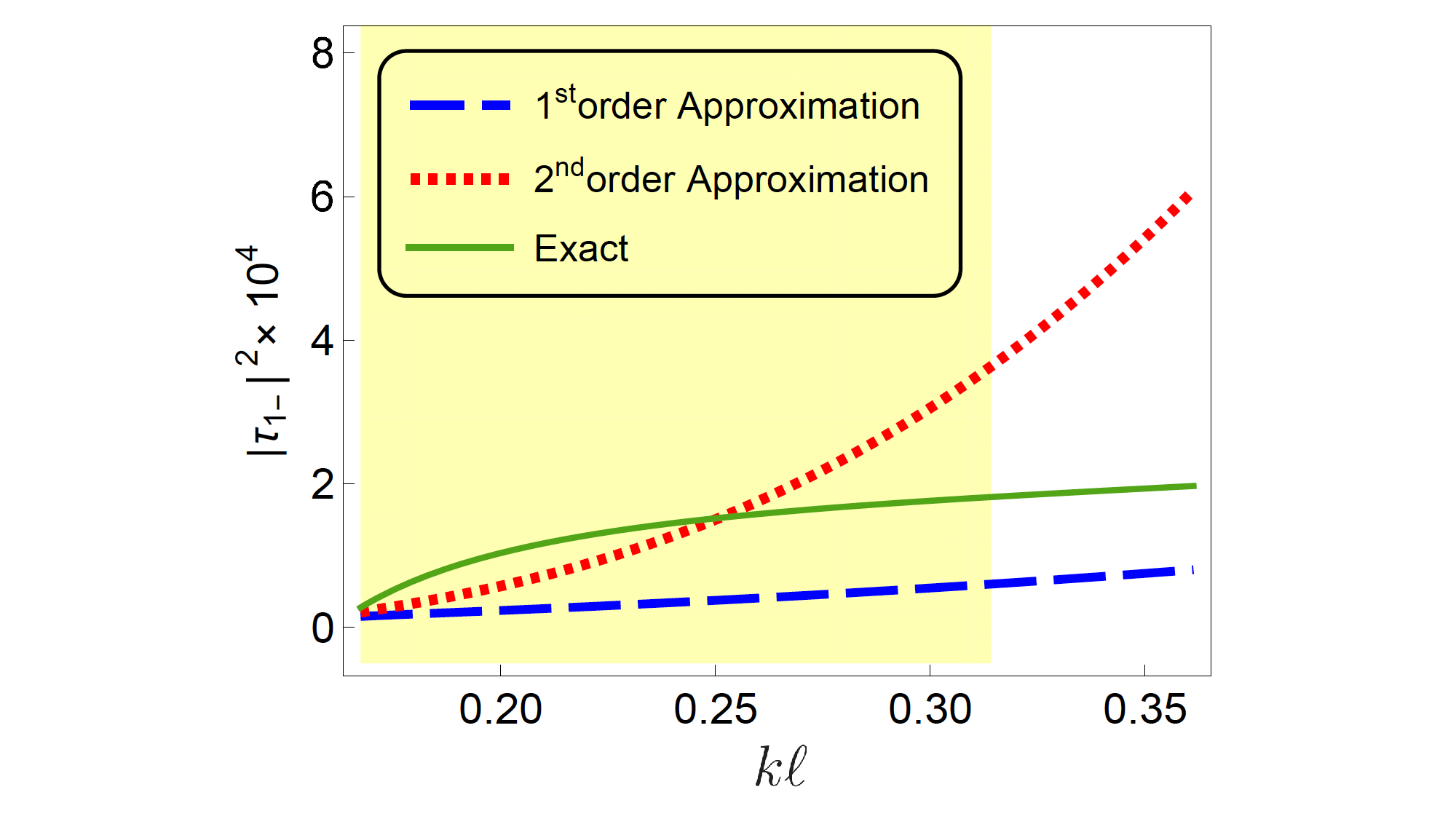}~\hspace{-1.5cm}~\includegraphics[scale=.3]{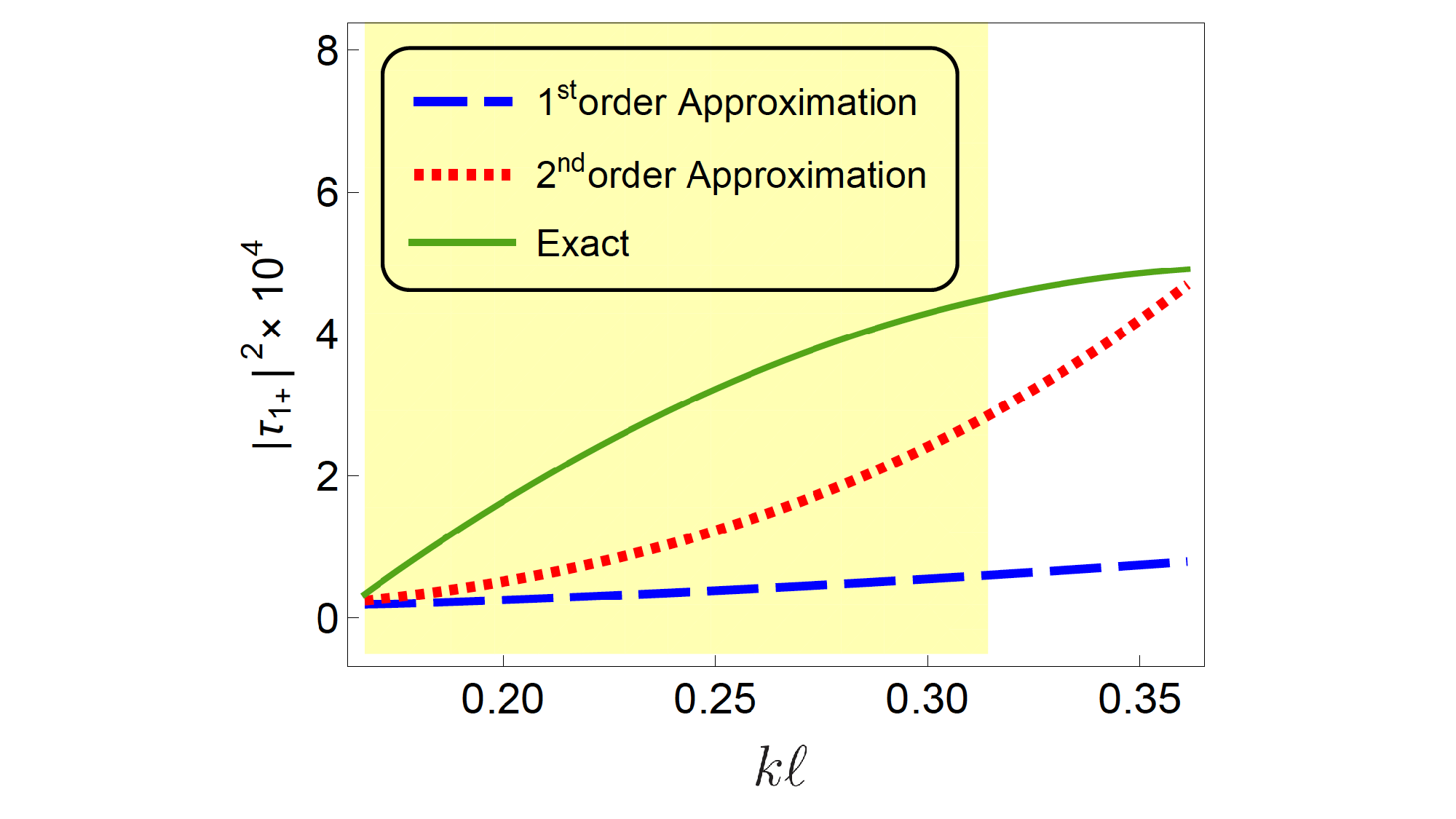}
       \caption{Graphs of $|\tau_{1\pm}|^2\times 10^4$ as functions of $k\ell$ for $\fz_1$ and $\fz_2$ given by \eqref{specs}, $\ell=100~{\rm nm}$, and $\fK=\pi~\mu{\rm m}^{-1}$. The yellow region corresponds to $\kappa_0<k\leq\fK$. The dashed and dotted curves respectively correspond to the outcome of the first- and second-order low-frequency approximations, while the solid curve corresponds to the exact calculation of $\tau_{0-}$.}
        \label{fig5}
        \end{center}
        \end{figure}

Figs.~\ref{fig4} and~\ref{fig5} show the graphs of the real and imaginary parts of $\tau_{0-}$ and those of $|\tau_{1\pm}|^2\times 10^{4}$ as functions of $k\ell$ for $\fz_0$ and $\fz_1$ given by \eqref{specs}, $\ell=100~{\rm nm}$, and $\fK=\pi~\mu{\rm m}^{-1}$. For these values of the parameters of the problem, $\kappa_0<k\leq\fK$ corresponds to $0.157<k\ell\leq 0.314$, and the wavelength of the incident wave ranges over $2$-$4~\mu{\rm m}$. {These figures also include the graphs corresponding to the exact calculation of $\tau_{0-}$ and $\tau_{1\pm}$ whose derivation we describe in Supplementary Materials \cite{SM}.} 

{As shown in Fig.~\ref{fig4}, the graphs of $\RE(\tau_{0-})$ and $\IM(\tau_{0-})$ obtained using the first- and second-order low-frequency approximations converge to the one corresponding to the exact calculation of these quantities when $k\ell$ tends to zero. The expected discrepancy between the approximate and exact results is small for first-order approximation provided that $k\ell<0.1$. It is negligible for the second-order approximation for $k\ell<0.2$. We reach similar conclusions when comparing the graphical representations of the approximate and exact calculations of $|\tau_{1\pm}|^2$. As shown in Fig.~\ref{fig5}, the values of  $|\tau_{1\pm}|^2$ obtained using the approximate and exact calculations converge to zero as $k\ell$ tends to $\kappa_0\ell=0.157$. For $k\ell<0.3$, the discrepancy between the approximate and exact results is smaller for the second-order approximation. For $k\ell<0.2$, the second-order approximation is highly reliable.}

\section{Low-frequency invisibility}
\label{S5}

%\subsection{Omnidirectional invisibility at low frequencies}

Substituting \eqref{WL-29} in \eqref{f1=}, we have
	\be
	\ff^{(1)}(\theta)=\frac{k}{2\sqrt{2\pi}}
	\Big[\fs_0 \fs\,\overline{\tilde v}_{\alpha,0}\big(k(\fs-\fs_0)\big)+
	\fc_0\fc \,
	\overline{\tilde w}_{\alpha,0}\big(k(\fs-\fs_0)\big)+
	\overline{\tilde w}_{\beta,0}\big(k(\fs-\fs_0)\big)\Big].
	\label{f1=explicit}
	\ee
This equation together with \eqref{series} show that the scattering amplitude $\ff(\theta)$ is proportional to $(k\ell)^2$ provided that 
	\be
	\overline{\tilde v}_{\alpha,0}(p)=\overline{\tilde w}_{\alpha,0}(p)=\overline{\tilde w}_{\beta,0}(p)=0~~\for~~p\in(-k,k).
	\label{condi-5-0}
	\ee
The media fulfilling this condition do not scatter TE and TM waves of sufficiently low frequencyies, i.e., they are effectively invisible for these waves.  

Let us use $\overline f(y)$ to denote $\int_0^1 d\cx\:f(\cx,y)$ for any function $f:[0,1]\times\R\to\C$ for which the latter exists as a function of $y$. Then, it is not difficult to see that \eqref{condi-5-0} holds, if 
	\begin{align}
	\overline{v}_{\alpha}(y)=\overline{w}_{\alpha}(y)=\overline{w}_{\beta}(y)=0,
	\label{condi-5-1}
	\end{align}
for all $y\in\R$. Recalling the definitions of $\alpha$, $\beta$, and $v_\alpha$, which are given by \eqref{alpha-beta} and \eqref{v-def}, and making use of \eqref{epsilon-L} -- \eqref{beta-L}, we can express \eqref{condi-5-1} in the form
	\begin{align}
	&\int_0^\ell\frac{dx}{\alpha(x,y)}=\int_0^\ell dx\: \hat\varepsilon(x,y)=\int_0^\ell dx\: \hat\mu(x,y)=\ell.
	\label{condi-5-2}
	\end{align}
	
Consider the cases where the medium is nonmagnetic, i.e., $\hat\mu=1$. Then, the last of equations \eqref{condi-5-2} holds identically. For a TE wave scattered by such a medium, $\alpha=\hat\mu=1$, and \eqref{condi-5-2} reduces to 
	\be
	\int_0^\ell dx\: \hat\varepsilon(x,y)=\ell.
	\label{condi-5-3}
	\ee
For a TM wave, $\alpha=\hat\varepsilon$, and \eqref{condi-5-2} is equivalent to the requirement that \eqref{condi-5-3} and
	\be
	\int_0^\ell \frac{dx}{\hat\varepsilon(x,y)}=\ell
	\label{condi-5-4}
	\ee
are satisfied. Expressing \eqref{condi-5-3} and \eqref{condi-5-4} in terms of the real and imaginary parts of $\hat\varepsilon$, we find
	\begin{align}
	&\int_0^{\ell} dx\: \RE[\hat\varepsilon(x,y)]=\int_0^{\ell} dx\: \frac{\RE[\hat\varepsilon(x,y)]}{|\hat\varepsilon(x,y)|^2}=\ell,
	\label{condi-5-3-real}\\
	&\int_0^{\ell} dx\: \IM[\hat\varepsilon(x,y)]=\int_0^{\ell} dx\: \frac{\IM[\hat\varepsilon(x,y)]}{|\hat\varepsilon(x,y)|^2}=0,
	\label{condi-5-4-imaginary}
	\end{align}
where `$\RE$' and `$\IM$' stand for the real and imaginary parts of their argument, respectively.

For a $\cP\cT$-symmetric nonmagnetic material satisfying $\hat\varepsilon(\ell-x,y)^*=\hat\varepsilon(x,y)$, 
	%\eqref{condi-5-3-real} takes the form
	%\begin{align}
	%&\int_0^{\ell/2} dx\: \RE[\hat\varepsilon(x,y)]=
	%\int_0^{\ell/2} dx\: \frac{\RE[\hat\varepsilon(x,y)]}{|\hat\varepsilon(x,y)|^2}
	%=\frac{\ell}{2},\nn
	%\end{align}
%while 
\eqref{condi-5-4-imaginary} holds automatically. Therefore, similarly to the one-dimensional setups \cite{pra-2013}, $\cP\cT$-symmetry facilitates  invisibility in two dimensions.

In the remainder of this section, we explore the application of the low-frequency invisibility conditions \eqref{condi-5-3} and \eqref{condi-5-4} to devise a cloaking scheme that prevents the scattering of low-frequency TE or TM waves of arbitrary incidence angle by a nonmagnetic slab ${\cS}_\star$ with translational symmetry along the $z$ axis. 

\begin{figure}
        \begin{center}
        \includegraphics[scale=.35]{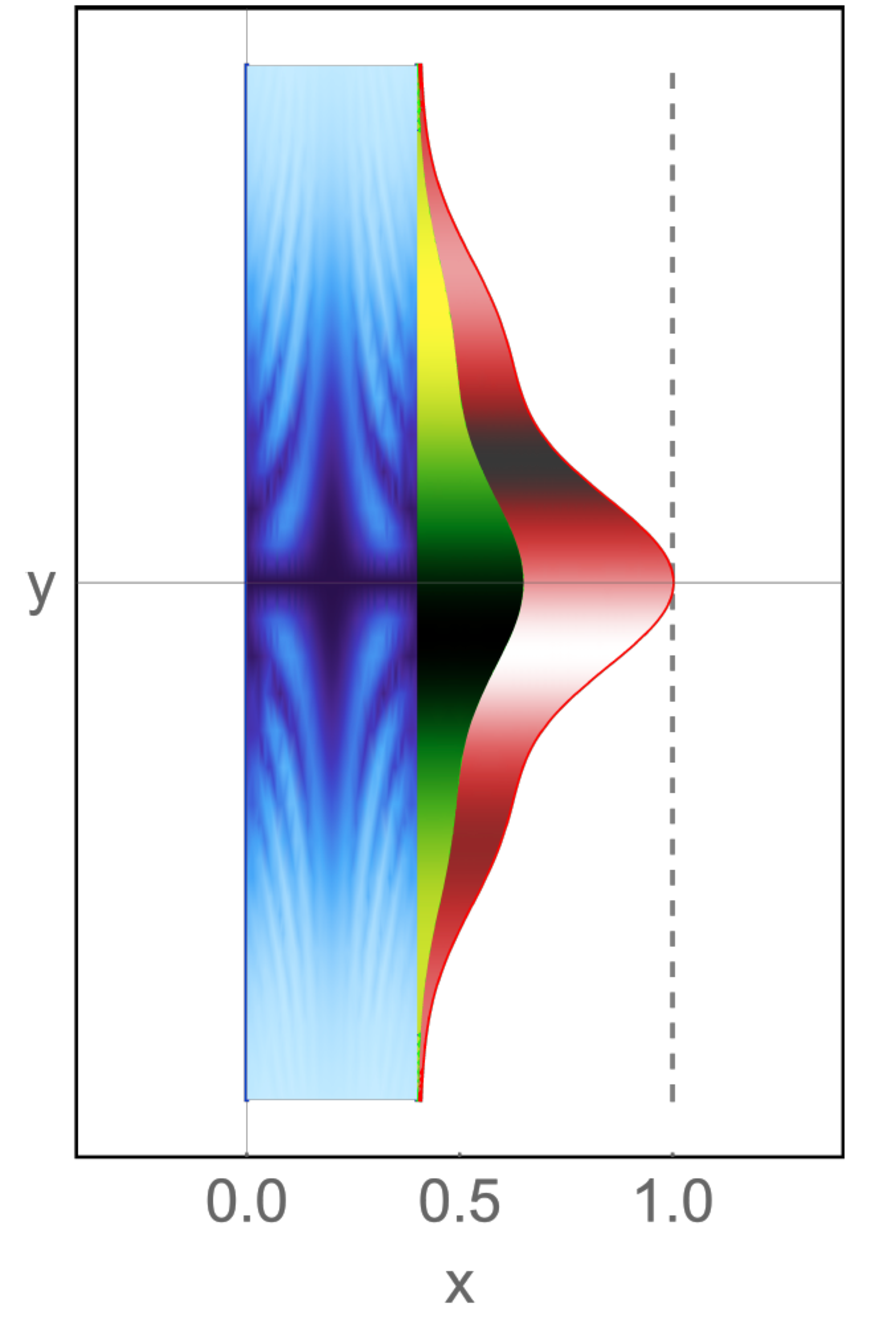}
        \caption{Schematic view of the slab $\cS_\star $ and its coatings $\cS_-$ and $\cS_+$ respectively colored in shades of blue, green, and red. {The coloring patterns represent the dependence of the inhomogeneities of the slab on both $x$ and $y$, and dependence of the inhomogeneities of the coatings on $y$.} Distances are measured in units of $\ell$.}
        \label{fig6}
        \end{center}
        \end{figure}

Consider coating ${\cS}_\star$ by a pair of layers $\cS_\pm$ of nonmagnetic dielectric materials or metamaterials that also possess translational symmetry along the $z$ axis. Suppose that the projections of $\cS_\star$, $\cS_-$, and $\cS_+$ on the $x$-$y$ plane are respectively given by
	\begin{align}
	&S_\star:=\left\{\left.(x,y)\in\R^2\:\right|\:0<x<\ell_\star\:\right\},\nn\\
	&S_-:=\left\{\left.(x,y)\in\R^2\:\right|\:\ell_\star<x<\ell_\star+\ell_-(y)\:\right\},\nn\\
	&S_+:=\left\{\left.(x,y)\in\R^2\:\right|\:\ell_\star+\ell_-(y)< x<
	\ell_{\cS}(y)\:\right\},\nn
	\end{align}
where $\ell_\star$ is the thickness of $\cS_\star$, $\ell_\pm:\R\to[0,\infty)$ are a pair of functions that determine the shapes of $\cS_\pm$, and $\ell_{\cS}:\R\to[0,\infty)$ is the function given by
	\be
	\ell_{\cS}(y):=\ell_\star+\ell_-(y)+\ell_+(y).
	\label{ell-s}
	\ee
{The latter} specifies the thickness of the coated slab, namely $\cS:=\cS_\star\cup\cS_-\cup\cS_+$. Clearly 
	\be
	\ell_\star\leq\ell_{\cS}(y)\leq\ell.
	\label{condi-coat}
	\ee	
Fig.~\ref{fig6} provides a schematic view of $S_\star$ and $S_\pm$.
        
Let $\hat\varepsilon_\star$ and $\hat\varepsilon_\pm$ respectively denote the relative permittivities of ${\cS}_\star$ and $\cS_\pm$, and suppose that $\hat\varepsilon_\pm$ depend only on $y$. Then the relative permittivity of the whole system takes the form
	\be
	\hat\varepsilon(x,y)=\left\{\begin{array}{cc}
	\hat\varepsilon_\star(x,y)&\for~~(x,y)\in  S_\star,\\
	\hat\varepsilon_-(y)&\for~~(x,y)\in  S_-,\\
	\hat\varepsilon_+(y)&\for~~(x,y)\in  S_+,\\
	1&{\rm otherwise}.\end{array}\right.
	\nn%\label{coated}
	\ee
Substituting this equation in \eqref{condi-5-3} and \eqref{condi-5-4}, we find the following conditions for the low-frequency invisibility of the system.
	\begin{align}
	&\hat\ell_-\hat\varepsilon_-+\hat\ell_+\hat\varepsilon_+=  u_+,
	&&\frac{\hat\ell_-}{\hat\varepsilon_-}+\frac{\hat\ell_+}{\hat\varepsilon_+}= u_-,
	\label{eq1-2}
	\end{align}
where we have suppressed the arguments of various functions for brevity and introduced
	\begin{align}
	&\hat\ell_\pm(y):=\frac{\ell_\pm(y)}{\ell_{\cS}(y)},
	&&u_{\pm}(y):=1-\hat\ell_\star(y)\sE_\pm(y)
	\label{up-um}\\
	&\hat\ell_\star(y):=\frac{\ell_\star}{\ell_{\cS}(y)},
	&&\sE_\pm(y):=\frac{1}{\ell_\star}\int_{0}^{\ell_\star}dx\:\hat\varepsilon_\star(x,y)^{\pm 1}.
	\label{sEpm-def}
	\end{align}
%and we have suppressed the $y$-dependence of $\hat\ell_\pm$, $\hat\varepsilon_\pm$, and $u_\pm$ for brevity.
%We can easily solve \eqref{eq1-2} for $\hat\varepsilon_\pm$. 

If  $\sE_-(y)=\sE_+(y)=1$, we can satisfy \eqref{eq1-2} by setting $\hat\ell_\pm(y)=0$. In this case, $\ell_\pm(y)=0$, $\ell_{\cS}(y)=\ell_\star$, and $u_\pm(y)=0$. If
	\be
	\sE_+(y)\neq 1~~{\rm or}~~\sE_-(y)\neq 1,
	\label{not-one}
	\ee 
$\hat\ell_-(y)+\hat\ell_+(y)>0$, $\hat\ell_{\cS}(y)>\ell_\star$, $\hat\ell_\star(y)<1$, and \eqref{eq1-2} admits the following general solution for $\ell_\pm(y)\neq 0$.
%	\begin{align}
%	&\hat\varepsilon_-=
%	\frac{u_-u_+- (\hat\ell_+^2-\hat\ell_-^2\pm u_0)}{2\hat\ell_- u_-},
%	&&\hat\varepsilon_+=
%	\frac{u_-u_++(\hat\ell_+^2-\hat\ell_-^2\pm u_0)}{2\hat\ell_+ u_-},
%	\label{sol-1}
%	\end{align}	
	\begin{align}
	&\hat\varepsilon_\pm=
	\frac{u_-u_+\pm (\hat\ell_+^2-\hat\ell_-^2+\varsigma\, u_0)}{2\hat\ell_\pm u_-},
	\label{sol-1}
	\end{align}
where $\varsigma$ is an indeterminate sign, and
	\begin{align}
	u_0&:=\sqrt{\big[u_- u_+-(\hat\ell_+-\hat\ell_-)^2\big]\big[u_- u_+-(\hat\ell_++\hat\ell_-)^2\big]}.
	\label{u-zero=}
	\end{align}
For $\ell_-=\ell_+$, \eqref{sol-1} and \eqref{u-zero=} respectively reduce to
%	\begin{align}
%	&\hat\varepsilon_-=
%	\frac{u_-u_+\mp u_0}{2\hat\ell_+ u_-},
%	&&\hat\varepsilon_+=
%	\frac{u_-u_+ \pm u_0}{2\hat\ell_+ u_-},
	%&&u_0:=\sqrt{u_-u_+\big(u_-u_+-4\hat\ell_+^2\big)}.
	%&&u_0(y):=u_-(y)u_+(y)\sqrt{1-\frac{4\hat\ell_-^2}{u_-(y)u_+(y)}}.
%	\label{sol-2}
%	\end{align}	
	\begin{align}
	&\hat\varepsilon_\pm=
	\frac{u_-u_+\pm\varsigma\, u_0}{2\hat\ell_+ u_-},
	\label{sol-2}
	\end{align}
and $u_0:=\sqrt{u_-u_+\big(u_-u_+-4\hat\ell_+^2\big)}$. Equation~\eqref{sol-2} shows that the solutions of \eqref{eq1-2}  given by the choice of opposite signs $\varsigma$ in \eqref{sol-2} correspond to configurations obtained by swapping the cloaking layers $\cS_\pm$. The same holds for the general solutions given by \eqref{sol-1}.

The solutions of \eqref{eq1-2} given by \eqref{sol-1} and  \eqref{sol-2} are admissible provided that $u_-(y)\neq 0$ and their right-hand sides are bounded away from zero, i.e., for all $y\in\R$, $|\hat\varepsilon_\pm(y)|$ be greater than some positive real number. The latter {condition} is necessary for the applicability of our general results on low-frequency scattering.

We can use \eqref{ell-s}, \eqref{up-um}, and \eqref{sEpm-def} to express $u_-(y)\neq 0$ in the form
	\be
	\ell_-(y)+\ell_+(y)\neq \ell_\star[\sE_-(y)-1],
	\nn
	\ee
which holds for generic choices of $\ell_\pm$.
	
In Appendix~B, we examine the consequences of the requirement that $\hat\varepsilon_\pm$ are bounded away from zero for situations where $\cS_\star$ is made of a typical lossless (and gainless) dielectric material. It turns out that for such material $u_-(y)\neq 0$, but $\hat\varepsilon_\pm$ do not take real and positive values for any choice of $\ell_\pm(y)$, i.e., the coating layers must involve regions of loss or gain, or be made of negative-index metamaterials. It is however possible to choose $\ell_\pm(y)$ in such a way that the real parts of $\hat\varepsilon_\pm(y)$ are positive. As we show in Appendix~B, this happens whenever 
	\begin{align}
	&\ell_{\rm min}(y)>\frac{1}{2}[\sE_-(y)+\sE_+(y)-2]\ell_\star,
	\label{condi-new36} \\
	&\ell_+(y)+\ell_-(y)>\frac{[\sE_+(y)-1][1-\sE_-(y)]\ell_\star^2}{2\ell_{\rm min}(y)+[2-\sE_-(y)-\sE_+(y)]\ell_\star}.
	\label{condi-new37} 
	\end{align}
where $\ell_{\rm min}(y)$ is the smallest of $\ell_+(y)$ and $\ell_-(y)$, i.e.,
$\ell_{\rm min}(y):=\ell_\mp(y)$ for $\ell_\pm(y)\geq\ell_\mp(y)$. Furthermore, for $\ell_-=\ell_+$, \eqref{condi-new36} and  \eqref{condi-new37} simplify considerably and reduce to
	\be
	\ell_-(y)=\ell_+(y)>\tfrac{1}{2}[\sE_+(y)-1]\ell_\star.
	\label{condi-new38} 
	\ee
Under this condition $u_0$ takes purely imaginary values. In view of \eqref{sol-2}, this implies that the real parts of $\hat\varepsilon_\pm$ coincide while their  imaginary parts have opposite signs. Furthermore, if we choose $\ell_+$ to be a constant multiple of $\sE_+-1$, then the real part of $\hat\varepsilon_+$ does not depend on $y$. This follows from \eqref{ell-s}, \eqref{up-um}, and \eqref{sol-2}.

As a concrete example, suppose that 
	\begin{align}
	&\hat\varepsilon_\star(x,y):=1+\fz\, e^{-\kappa x} e^{-\frac{y^2}{2{L}^2}},
	&&\ell_\pm(y):=\alpha\,\ell_\star e^{-\frac{y^2}{2L^2}},
	\label{eg}
	\end{align}
where $\fz$, $\kappa$, ${L}$, and $\alpha$ are positive real parameters. Then,
	\begin{align}
	&\sE_+(y)=1+\frac{\fz\, e^{-\frac{y^2}{2L^2}}\big(1-e^{-\kappa\ell_\star}\big)}{\kappa\ell_\star},
	%\label{sEp-eg}\\
	&&\sE_-(y)=1-\frac{1}{\kappa\ell_\star}\ln\left(
	\frac{1+\fz\, e^{-\frac{y^2}{2L^2}}}{1+\fz\, e^{-\kappa\ell_\star}e^{-\frac{y^2}{2L^2}}}\right).
	\label{sEpm-eg}
	\end{align}
For $\alpha>\fz/2\kappa\ell_\star$, this choice for $\ell_\pm$ satisfies \eqref{condi-new38}. Using  \eqref{ell-s}, \eqref{up-um},  \eqref{sEpm-def}, \eqref{sol-2}, and \eqref{sEpm-eg}, we find
	\begin{align}
	&\RE(\hat\varepsilon_\pm)=\rho:=1-\frac{\fz(1-e^{-\kappa\ell_\star})}{2\alpha\kappa\ell_\star},
	&&\IM[\hat\varepsilon_\pm(y)]=\pm\,\varsigma\,\rho
	\sqrt{\frac{1}{\rho\,\{\frac{\ell_\star[1-\sE_-(y)]}{2\ell_+(y)}+1\}}-1}.
	\end{align}
For $\varsigma=-$, we have $\mp\,\IM[\hat\varepsilon_\pm(y)]>0$. This means that the outer coating layer, namely $\cS_+$, has gain while $\cS_-$ is lossy \cite{silfvast}.	
	
\begin{figure}
        \begin{center}
        \includegraphics[scale=.36]{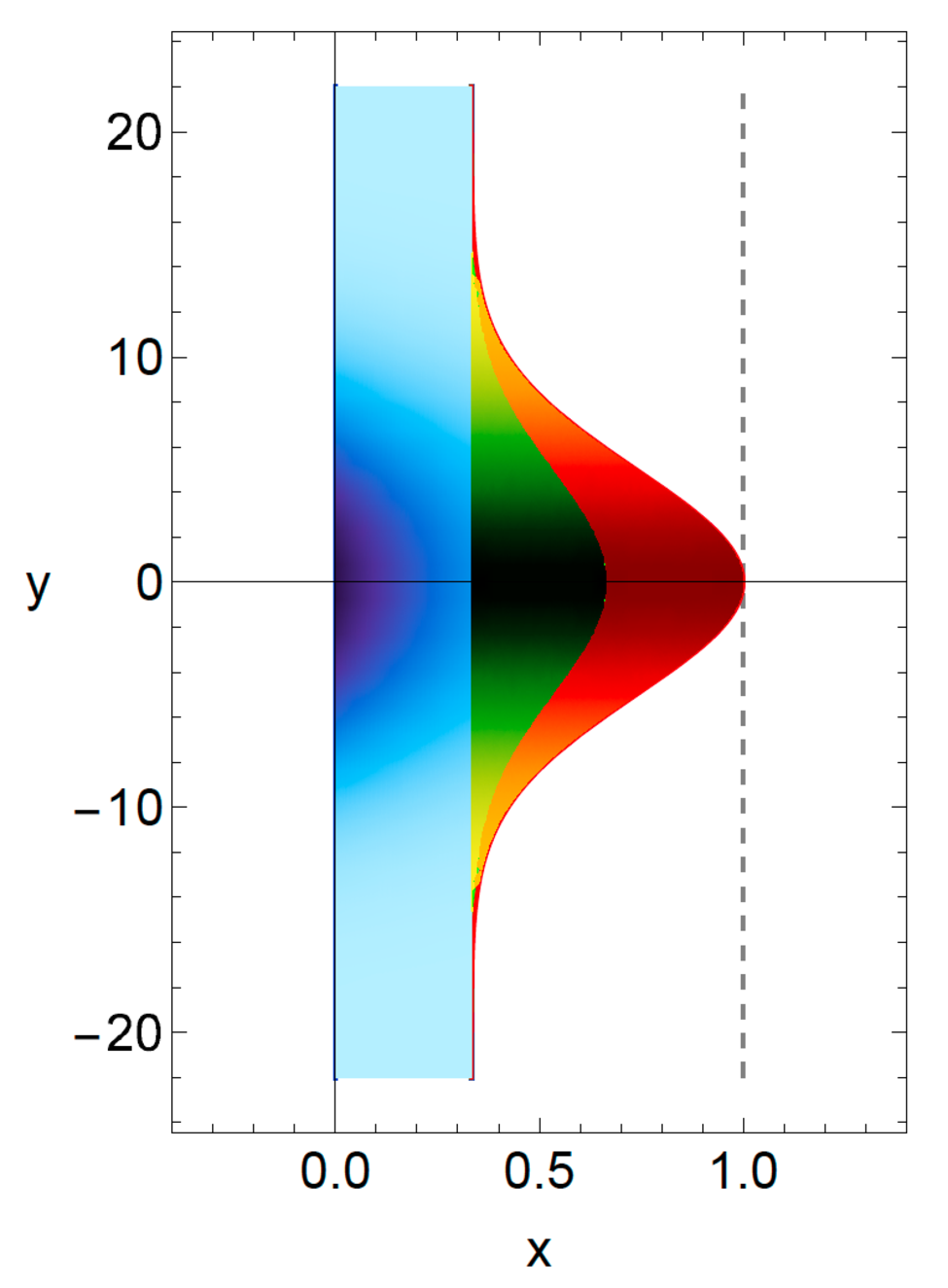}~~~~~~~~~~
        \includegraphics[scale=.77]{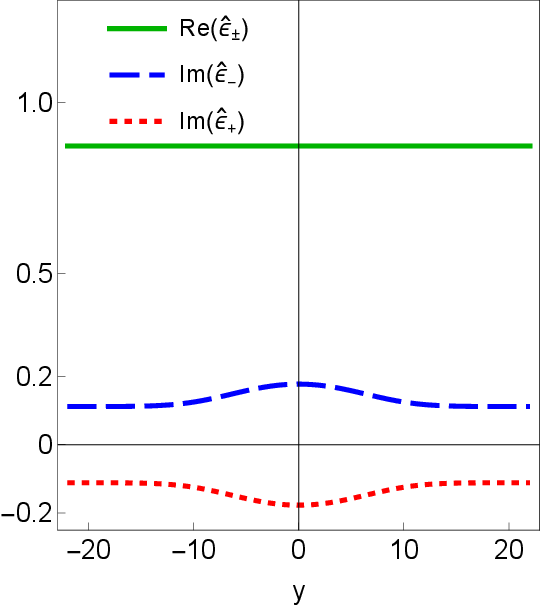}
        \caption{Schematic view of the coated slab on the left and plots of the real and imaginary parts of $\hat\varepsilon_\pm$ as a function of $y$ for $\hat\varepsilon_\star$ given by \eqref{eg} and \eqref{spec3}. The regions in the $x$-$y$ plane painted in shades of blue, green, and red respectively represent $S_\star$, $S_-$, and $S_+$. {The shadings correspond to density plots of $|\hat\varepsilon|$.} The distances are measured in units of $\ell$.}
        \label{fig7}
        \end{center}
        \end{figure}

Figure~\ref{fig7} provides a schematic representation of the coated slab and plots of  the real and imaginary parts of $\hat\varepsilon_\pm(y)$ for
	\begin{align}
	&\varsigma=-, &&\fz=0.4, &&\kappa=3\ell^{-1}, && L=5\ell, &&\ell_\star=\ell/3, &&\alpha=1.
	\label{spec3}
	\end{align}
These imply $\fz/2\kappa\ell_\star=0.2<\alpha$ and $\RE(\hat\varepsilon_\pm)=0.87$.

\section{Summary and Conclusions}
\label{S6}

Scattering has been an important subject of research in physics and mathematics since the 19th century. The progress made during the 20th century toward a systematic description of this phenomenon has led to a well-developed body of knowledge known as scattering theory. The standard textbook treatment of this theory makes use of Green's functions (resolvent operators), the scattering matrix, and the Lippmann-Schwinger equation as its basic tools \cite{yafaev,newton-book,taylor}. 

Recently, we have developed an alternative approach to wave scattering that relies on extracting scattering data from the time-evolution operator for an effective non-unitary quantum system \cite{ap-2014,pra-2021,pra-2023}. This provides a dynamical formulation of stationary scattering in which the role of the scattering matrix is played by a fundamental notion of transfer matrix. The latter is an integral operator acting in an infinite-dimensional function space. The numerical transfer matrices employed in the study of wave propagation and scattering in the 1980's and 1990's \cite{pendry-1984,pendry-1990a,pendry-1994} correspond to the approximations of this operator by large matrices. 

An important advantage of the fundamental transfer matrix over the numerical transfer matrices is that it allows for analytic treatment of various scattering problems. Among the most notable of its applications is the recent discovery of the scattering potentials and permittivity/permeability profiles whose scattering problems are exactly solvable by the Born approximation of order $N$ with $N$ depending on the incident wavenumber \cite{ptep-2024a,jpa-2024}. Another appealing feature of this approach is its effectiveness in performing analytic low-frequency scattering calculations for scalar waves satisfying the Schr\"odinger or Helmholtz equations \cite{jmp-2021,jpa-2021,pra-2025}. 

In the present article, we have developed a dynamical formulation of the stationary scattering of TE and TM waves by the inhomogeneities of an effectively two-dimensional medium. These waves are described by the two-dimensional Bergmann equation whose treatment cannot be directly mapped to the Helmholtz equation for typical realistic setups. The technical difficulties associated with the scattering of these waves do not allow for a direct extension of the known results in one dimension \cite{ptep-2024b,jpa-2025} either. To deal with the scattering of these waves, we have therefore constructed an appropriate effective Hamiltonian operator and showed how its evolution operator determines the scattering amplitude. 

A remarkable property of this Hamiltonian operator is that its products, which determine the Dyson series for its evolution operator, admit a general expression whose terms can be iteratively calculated. This allows for developing a systematic method of constructing the low-frequency series for the system's scattering amplitude. Using this method, we have found explicit analytic expressions for the first- and second-order terms in this series and checked the validity of the related low-frequency approximations by comparing their outcomes with the exact results for a class of exactly solvable grating systems. We have further used our general results to obtain a quantitative characterization of low-frequency invisibility and to devise a cloaking scheme that suppresses the scattering of low-frequency TE and TM waves of arbitrary incidence angle by an effectively two-dimensional slab.

Since the scattering of sound waves is also described by Bergmann's equation, our results are readily applicable for acoustic scattering in two dimensions. As a subject for future study, we plan to extend these results to acoustic scattering in three dimensions. \vspace{12pt}

\subsection*{Acknowledgements} 
This work has been supported by the Scientific and Technological Research Council of T\"urkiye (T\"UB\.{I}TAK) in the framework of the project 123F180 and by Turkish Academy of Sciences (T\"UBA).

\section*{Appendix~A: Derivations of Eqs.~\eqref{f1=} and \eqref{f2=}} 

To obtain explicit expressions for the leading- and next-to-leading-order terms in the low-frequency expansion of the scattering amplitude we need to compute $\widehat N_{ab}^{(1)}$ and $\widehat N_{ab}^{(2)}$. In view of \eqref{Nab-m=2}, this requires the knowledge of $\widehat\bsL_1^{\,(0)}$, $\widehat\bsL_1^{\,(1)}$, and $\widehat\bsL_2^{\,(0)}$. We can use \eqref{VQ-def}, \eqref{bL-def}, \eqref{P-def}, and \eqref{bcL-def} -- \eqref{bcL-series} to show that
	\begin{align}
	&\widehat\bsL_1^{\,(0)}=\widehat\Pi_k \left[\begin{array}{cc}
	\widehat 0 & \widehat{\overline \sW}_0 \\
	\widehat{\overline w}_{\alpha,0} & \widehat 0\end{array}\right]\widehat\Pi_k,
	\quad\quad\quad
	\widehat\bsL_2^{\,(0)}=\widehat\Pi_k\left[\begin{array}{cc}
	\widehat X& \widehat 0\\
	\widehat 0&\widehat Y \end{array}\right]\widehat\Pi_k,
	\label{bcL0}\\
	&\widehat\bsL_1^{\,(1)}=i\widehat\Pi_k \left[\begin{array}{cc}
	\widehat{\overline \sW}_1-\widehat{\check\varpi}^2\widehat{\overline w}_{\alpha,1}  &\widehat 0\\
	\widehat 0 &-\widehat{\overline \sW}_1+\widehat{\overline w}_{\alpha,1} \widehat{\check\varpi}^2 \end{array}\right]\widehat\Pi_k,
	\label{bcL11}
	\end{align}
where 
	\begin{align}
	&\widehat{\overline \sW}_l:=\int_0^1d\cx\:\cx^l\,\widehat{\check\sW}(\cx)=
	k^{-2}\widehat p~ \widehat{\overline{v}}_{\alpha, l}\: \widehat p+
	\widehat{\overline w}_{\beta,l},\quad\quad
	\widehat{\overline v}_{\alpha,l} :=\int_0^1d\cx\:\cx^l\,\widehat{v}_\alpha(\cx),\\
	&\widehat{\overline w}_{\alpha,l} :=\int_0^1d\cx\:\cx^l\,\widehat{w}_\alpha(\cx), 
	\qquad\qquad\qquad~~~
	\widehat{\overline w}_{\beta,l} :=\int_0^1d\cx\:\cx^l\,\widehat{w}_\beta(\cx),\\
	&\widehat X:=\int_0^1d\cx_2\!\int_0^{\cx_2}\!\!d\cx_1\:\widehat{\check\sW}(\cx_2)\widehat{w}_\alpha(\cx_1),
	\quad\quad
	\widehat Y:=\int_0^1d\cx_2\!\int_0^{\cx_2}\!\!d\cx_1\:
	\widehat{w}_\alpha(\cx_2)\widehat{\check\sW}(\cx_1),
	\end{align}
and $l\in\{0,1\}$. Equations \eqref{Nab-m=2}, \eqref{bcL0}, and \eqref{bcL11} imply
	\begin{align}
	\widehat N_{ab}^{(1)}=&\frac{i}{2}\widehat\Pi_k\Big[(-1)^a\widehat{\overline \sW}_0\,
	\widehat{\check\varpi}^{_{-1}}
	+(-1)^b\widehat{\check\varpi}\:\widehat{w}_{\alpha,0}\Big]\widehat\Pi_k,
	\label{Nab-1n}\\
	\widehat N_{ab}^{(2)}=&\frac{1}{2}\widehat\Pi_k\Big[(-1)^{a+b}
	(\widehat X+\widehat{\overline \sW}_1-\widehat{\check\varpi}^2
	\widehat{w}_{\alpha,1})
	 +\widehat{\check\varpi}(\widehat Y-\widehat{\overline \sW}_1+
	 \widehat{w}_{\alpha,1}\widehat{\check\varpi}^2)\widehat{\check\varpi}^{_{-1}}\Big]
	 \widehat\Pi_k.
	\label{Nab-2n}
	\end{align}

Next, we note that given a function $f:\R^2\to\C$ and $p,p'\in\R$,
	\be 
	\br p|f(\cx,\widehat y)|p'\kt=\frac{1}{2\pi}\int_{-\infty}^\infty dy\:e^{-i(p-p')y}f(\cx,y)=
	\frac{\tilde f(\cx,p-p')}{2\pi}.
	\label{pfp}
	\ee
Using  \eqref{Nab-1n} -- \eqref{pfp}, we can show that 
	\begin{align}
	\br p|\widehat N_{ab}^{(1)}|p'\kt=&\;\frac{i\chi_k(p)\chi_k(p')}{4\pi}	
	\Bigg\{\frac{(-1)^a\, W_0(p,p')}{\sqrt{1-p^{\prime 2}/k^2}}
	+(-1)^b \sqrt{1-p^{2}/k^2}\:\overline{\tilde w}_{\alpha,0}(p-p')\Bigg\},
	\label{Nab-1-ME}\\
	\br p|\widehat N_{ab}^{(2)}|p'\kt=&\;\frac{\chi_k(p)\chi_k(p')}{4\pi}	
	\Bigg\{ (-1)^{a+b}
	\Big[ X_1(p,p')+ X_2(p-p')+ W_1(p,p')\nn\\
	&\hspace{2.6cm}-\mbox{$\left(1-\frac{p^2}{k^2}\right)$}\overline{\tilde w}_{\alpha,1}(p-p')\Big]
	+\mbox{$\sqrt{\frac{k^2-p^2}{k^2-p^{\prime2}}}$}\Big[	 
	  Y_1(p,p')+\nn\\
	 &\hspace{2.8cm}Y_2(p-p')- W_1(p,p')+\mbox{$\left(1-\frac{p^{\prime 2}}{k^2}\right)$}
	 \overline{\tilde w}_{\alpha,1}(p-p')\Big]\Bigg\},
	 \label{Nab-2-ME}
	\end{align}
where 
	\[\chi_k(p):=\left\{\begin{array}{cc}
	1 &\for~~|p|<k,\\
	0 &\for~~|p|\geq k,\end{array}\right.\]
and $W_l$ and $\overline{\tilde w}_{\alpha,l}$ with $l\in\{0,1\}$ and $X_j$, and $Y_j$  with $j\in\{1,2\}$ are given by \eqref{W-def} -- \eqref{Y2-def}.
	
Substituting \eqref{Nab-1-ME} and \eqref{Nab-2-ME} in the second equation in \eqref{M-expand} and using the resulting expression together with \eqref{f=}, \eqref{B-L-series} -- \eqref{A-R-series}, and \eqref{series}, we obtain \eqref{f1=} and \eqref{f2=}.

\section*{Appendix~B: $\hat\varepsilon_\pm$ for a slab made of typical lossless dielectric material} 

Suppose that $\cS_\star$ is made of a typical lossless (and gainless) dielectric material, so that $\hat\varepsilon_\star$ is a piecewise-continuous real-valued function, and $\hat\varepsilon_\star(x,y)\geq 1$ for all $(x,y)\in\R^2$. 
Then $\sE_\pm$ and $u_\pm$ are real-valued functions, and \eqref{sEpm-def} and \eqref{not-one} imply
	\begin{align}
	&0<\sE_-(y)<1<\sE_+(y),
	\label{sEmp-condi-1}\\
	&\sE_-(y)+\sE_+(y)>2.
	\label{sEmp-condi-2}
	\end{align}
These relations together with \eqref{ell-s}, \eqref{up-um}, \eqref{sEpm-def}, and $0<\hat\ell_\star(y)<1$ allow use to establish
	\begin{align}
	&u_-(y)>0,
	\label{condi-new1}
	\end{align}
and 
	\begin{align}
	u_-u_+&=\ell_\cS^{-2}(\ell_\cS-\ell_\star\sE_+)(\ell_\cS-\ell_\star\sE_-)\nn\\
	&=\ell_\cS^{-2}[(\ell_-+\ell_++\ell_\star)^2-\ell_\star(\ell_-+\ell_++\ell_\star)(\sE_++\sE_-)+\ell_\star^2\sE_-\sE_+]\nn\\
	&=\ell_\cS^{-2}[(\ell_-+\ell_+)^2+\ell_\star(\ell_-+\ell_+)(2-\sE_--\sE_+)+
	\ell_\star^2(1-\sE_+)(1-\sE_-)]\nn\\
	&=(\hat\ell_-+\hat\ell_+)^2+\hat\ell_\star(\hat\ell_-+\hat\ell_+)(2-\sE_--\sE_+)+
	\hat\ell_\star^2(1-\sE_+)(1-\sE_-).
	\label{uu-calc}
	\end{align}
In view of \eqref{sEmp-condi-1} and \eqref{sEmp-condi-2},
the terms proportional to $\hat\ell_\star$ and $\hat\ell_\star^2$ on the right-hand side of \eqref{uu-calc} take negative real values. This, in particular, implies that
	\be
	u_-(y)u_+(y)<[\hat\ell_-(y)+\hat\ell_+(y)]^2.
	\label{condi-new2}
	\ee
Similarly, with the help of \eqref{ell-s}, \eqref{up-um}, and $0<\hat\ell_\star(y)<1$, we can show that $u_+(y)>0$ if and only if
	\be
	\ell_-(y)+\ell_+(y)>[\sE_+(y)-1]\ell_\star.
	\label{up-positive}
	\ee

Next, we note that the necessary and sufficient condition for the $\hat\varepsilon_\pm$ given by \eqref{sol-1} to take positive real values is that the same holds for $u_0$ and $u_+$, and 
%	\be
%	u_-(y)u_+(y)>|\hat\ell_+(y)^2-\hat\ell_-(y)^2\pm u_0(y)|.
%	\nn%\label{condi-new0}
%	\ee
%By virtue of the requirement that $u_0(y)>0$, this is equivalent to 
	 \be
	u_-(y)u_+(y)>|\hat\ell_+(y)^2-\hat\ell_-(y)^2|+ u_0(y).
	\label{condi-upum}
	\ee
	
According to \eqref{u-zero=} and \eqref{condi-new1}, if $u_0$ is real-valued and $u_+(y)>0$, we have $u_0(y)>0$ and $u_-(y)u_+(y)>0$. The latter condition and \eqref{condi-new2} imply
	\be
	u_-(y)u_+(y)<[\hat\ell_-(y)-\hat\ell_+(y)]^2.
	\nn
	\ee
Combining this inequality and \eqref{condi-upum}, we obtain $u_0(y)<\Delta(y)$, where
	\begin{align}
	\Delta(y)&:=[\hat\ell_-(y)-\hat\ell_+(y)]^2-|\hat\ell_-(y)^2-\hat\ell_+(y)^2|\nn\\
	&=-2\,\ell_\mp(y)|\ell_-(y)-\ell_+(y)|~~\for~~\ell_\pm(y)\geq\ell_\mp(y)\nn\\
	&\leq 0.
	\label{Delta-1}
	\end{align}
This gives $u_0(y)<0$, which contradicts the requirement that $u_0(y)>0$. Therefore, conditions $u_0(y)>0$, $u_+(y)>0$, and \eqref{condi-upum} are inconsistent. This shows that $\hat\epsilon_\pm(y)$ cannot be both real and positive.  

We can however use \eqref{sol-1} to obtain configurations in which the real parts of $\hat\varepsilon_\pm(y)$ are positive. To see this, suppose that $\hat\ell_\pm(y)$ satisfy
	\be
	u_-(y)u_+(y)>|\hat\ell_-(y)^2-\hat\ell_+(y)^2|.
	\label{condi-new10}
	\ee
Then, in view of \eqref{condi-new2} and \eqref{Delta-1}, we have
	\[ [\hat\ell_-(y)-\hat\ell_+(y)]^2<u_-(y)u_+(y)<[\hat\ell_-(y)+\hat\ell_+(y)]^2.\]
Using this relation and \eqref{u-zero=}, we see that $u_0$ takes imaginary values. This observation together with \eqref{condi-new1} show that whenever \eqref{condi-new10} holds,
	\begin{align}
	&\RE[\hat\varepsilon_\pm(y)]=\frac{u_-(y)u_+(y)\pm[\hat\ell_+(y)^2-\hat\ell_-(y)^2)]}{2\,\hat\ell_\pm (y)u_-(y)}>0,\nn\\
	&\IM[\hat\varepsilon_\pm(y)]= \frac{\pm\,\varsigma\: |u_0(y)|}{2\,\hat\ell_\pm (y)u_-(y)}.\nn
	\end{align}
Note that for $\ell_-=\ell_+$, $\RE(\hat\varepsilon_\pm)=u_+/2\hat\ell_+$. In particular, if we choose $\ell_+$ to be a constant  multiple of $u_+$, then the real part of $\RE(\hat\varepsilon_\pm)$ will not depend on $y$.
	
Finally, we wish to point out that, in view of \eqref{ell-s}, \eqref{up-um},
\eqref{sEmp-condi-1} and \eqref{sEmp-condi-2}, \eqref{condi-new10} is equivalent to imposing conditions \eqref{condi-new36} and \eqref{condi-new37} on $\ell_\pm$. For $\ell_-=\ell_+$, we can respectively express these conditions in the form 
	\begin{align}
	&\ell_+(y)-[A_-(y)+A_+(y)]>0,
	&&[\ell_+(y)-A_-(y)][\ell_+(y)-A_+(y)]>0,
	\end{align}
where $A_\pm(y):=\frac{1}{2}[\sE_\pm(y)-1]\ell_\star$. Because $A_-(y)<0<A_+(y)$ and $\ell_+(y)>0$, these two inequalities are equivalent to $\ell_+(y)>A_+(y)$ which is the identical to \eqref{condi-new38} and \eqref{up-positive} with $\ell_-=\ell_+$. We arrive at the same conclusion by noting that because $u_-(x)>0$, for $\ell_-=\ell_+$, \eqref{condi-new10} is equivalent to $u_+(y)>0$ and consequently \eqref{up-positive}.

\ed